# Stability of buoyant-Couette flow in a vertical porous slot

B.M. Shankar[1], I.S. Shivakumara[2]

[1]Department of Mathematics, PES University, Bangalore 560 085, India.
[2]Department of Mathematics, Bangalore University, Bangalore 560 056, India.
**Corresponding Author**: B.M. Shankar; email: bmshankar@pes.edu

Abstract

The stability of two-dimensional buoyancy-driven convection in a vertical porous slot, wherein a plane Couette flow is additionally present, is studied. This complex fluid flow scenario is examined under the influence of Robin-type boundary conditions, which are applied to perturbations in both velocity and temperature. The inclusion of a time-derivative velocity term within the Darcy momentum equation notably introduces intricacies to the study. The stability of the basic natural convection flow is primarily governed by several key parameters namely, the Péclet number, the Prandtl-Darcy number, the Biot number and a non-negative parameter that dictates the nature of the vertical boundaries. Through numerical analysis, the stability eigenvalue problem is solved for a variety of combinations of boundary conditions. The outcomes of this analysis reveal the critical threshold values that signify the onset of instability. Furthermore, a detailed examination of the stability of the system has provided insights into both its commonalities and distinctions under different conditions. It is observed that, except for the scenario featuring impermeable-isothermal boundaries, the underlying base flow exhibits instability when subjected to various other configurations of perturbed velocity and temperature boundary conditions. This underscores the notion that the presence of Couette flow alone does not suffice to induce instability within the system. The plots depicting neutral stability curves show either bi-modal or uni-modal characteristics, contingent upon specific parameter values that influence the onset of instability.

1. Introduction

The convective instability in a differentially heated vertical porous layer due to the buoyancy force is a topic that has garnered great interest over the last few decades. There are several areas of engineering and physics within which such investigations are applied, for example, thermal insulation of buildings, devices such as breathing walls, geophysics, and the design of filtration systems. The pioneering work by Gill [1] on the conduction regime in a vertical porous layer bounded by impermeable isothermal boundaries subjected to a horizontal temperature gradient provided a simple and rigorous proof that the basic flow is unconditionally stable for all infinitesimal perturbations. Subsequently, this work laid the groundwork for numerous studies, which have either corroborated or modified the conclusions of Gill's classical proof. Several effects were introduced in these studies, such as nonlinear analysis of the problem (Straughan [2]; Flavin and Rionero [3]), inclusion of the time-derivative of the velocity term in the momentum balance equation (Rees [4]), consideration of local thermal nonequilibrium within the saturated porous material (Rees [5]; Scott and Straughan [6]), and exploration of power-law fluid saturating a porous layer (Barletta and Alves [7]). All of these studies have upheld the original finding that convective instability is not possible in the system. While changing impermeable boundaries to permeable ones (Barletta [8]), addition of boundary and inertia effects (Shankar et al. [9]), maximum density with finite Darcy-Prandtl number (Naveen et al. [10]), second diffusing component (Shankar et al. [11]), heterogeneity in permeability (Shankar and Shivakumara [12]), a three-layer porous slab with impermeable isothermal boundaries (Barletta et al. [13]) and variable viscosity fluid (Shankar et al. [14]) have contributed to altering the conclusion of Gill [1]. In a separate development, Barletta et al. [15] investigated the instability of buoyant flow in a vertical cylindrical porous slab with permeable boundaries and revealed that the basic parallel flow becomes even more unstable as the aspect ratio increases. By employing a numerical solution method, the validity of Gill's theorem was shown to be true for the case of an annular porous layer whose boundaries are kept at different uniform temperatures and they are impermeable. By relaxing the impermeability condition with a partial permeability through Robin boundary conditions for the pressure, they further disclosed that an instability emerges which takes the form of axisymmetric normal modes.

There is a different class of flow which is due to the movement of plates moving in opposite directions, called the plane Couette flow. Rayleigh [16] suggested that the plane Couette flow is stable for all infinitesimal disturbances but may be unstable for finite disturbances, which is a cornerstone result of fluid mechanics. The initial comprehensive demonstration of stability appears to be credited to Romanov [17]. He provided proof that the normal modes of the linear problem experience damping for all combinations of Reynolds and wave numbers. Since then, this field has continuously drawn the interest of researchers, and the chronological advancements are thoroughly documented in the book by Drazin and Reid [18]. In spite of theoretical outcomes derived from the linear stability analysis, various experiments, including those documented by Bottin et al. [19] and Tilmark and Alfredsson [20], have observed the occurrence of instability and turbulence as a response to finite-amplitude perturbations. These studies have successfully confirmed the transition to instability at Reynolds number approximately equal to 300. Recently, the plane Couette flow problem was extended to the porous domain by Shankar et al. [21] and showed the possibility of occurring unstable mode of disturbances prompting stabilizing and destabilizing effects on the base flow. The stability characteristics of an inviscid fluid and of the Darcy flow are also analyzed and found that the basic flow remains unconditionally stable to small-amplitude disturbances.

The buoyancy-driven motion of a fluid heated from the side (vertical layer) or below (horizontal layer), and plane Couette flow, are paradigms in the physical and engineering sciences and have been studied extensively to gain insights into turbulence in the literature. There are profound differences between these two canonical problems as well. Another area of research that has received considerable attention is the mixed convection problem. This intriguing phenomenon entails the combined effects of plate movement in opposite directions and buoyancy forces triggered by a constant temperature difference. Most of the works carried out till date correspond to the Couette flow in a horizontal fluid layer which are maintained at different temperatures, commonly referred to as Rayleigh-Bénard-Couette convection (Gallagher and Mercer [22]; Deardorff [23]; Ingersoll [24]). Of late, there has been a renewed interest in Rayleigh-Bénard-Couette systems as prototype problems to understand viscous dissipation induced thermal instabilities (Barletta and Nield [25]), transient growth phenomenon (Jerome et al. [26]) as well as the influence of thermal stratification on transient growth (Hu et al. [27]). A temporal stability analysis was performed for the Rayleigh-Bénard-Couette convection of viscoelastic fluids using the Oldroyd-B model in various point of view by Alves et al. [28]. In a nuclear power plant, if the circulation of the coolant is disrupted, a gap may form between a control rod and the reactor core. This situation results in heat transfer from the reactor core to the control rod, driven by the temperature difference between the two components, where the control rod generally has a lower temperature than the reactor core. To simulate and study this scenario, a vertical porous slot with a large aspect ratio can be employed as a suitable model. The Couette flow component is undoubtedly superimposed, albeit transiently, during the insertion of the control rods into or removal from the reactor core. Besides, the imposition of a Couette-like condition on the boundaries of a vertical porous layer is not only feasible but also highly relevant in numerous practical applications across different disciplines, including geophysics, engineering, environmental science, and materials science. The effect of plane Couette flow on linear and nonlinear stability of natural convection in an infinitely tall fluid layer is investigated by Tsunoda and Fujimura [29]. However, its counterpart in the porous medium has not been given any attention in the literature.

The primary objective of this paper is to offer a novel perspective on the stability analysis of natural convection within a fluid-saturated vertical porous layer accounting the presence of an overlaying plane Couette flow and the impact of the Prandtl-Darcy number. This paper exclusively focuses on the analysis of two-dimensional disturbances, given the evident validity of Squire's theorem (Rees [4]; Shankar et al. [30]). The investigation considers first-kind conditions in obtaining the solution for the basic state. Whereas Robin boundary conditions are applied to perturbations in both velocity and temperature facilitating the effects of diverse boundary condition combinations on the stability of the underlying base flow. The examination encompasses scenarios involving finite and infinite values of the Prandtl-Darcy number. This work holds significance due to the profound influence of boundary characteristics on flow stability, a factor that holds relevance in numerous practical applications, as documented in previous works (Barletta [8]; Barletta et al. [31]). By numerically solving the linear stability eigenvalue problem, the study discerns the sensitivity of governing parameters in either stabilizing or destabilizing the basic flow. Remarkably distinct results emerge for distinct boundary configurations. Furthermore, this investigation establishes connections with the findings of Rees [4] and Barletta [8] as specific instances, thereby augmenting the comprehensiveness of understanding surrounding the problem and its wider implications.

## 2. Statement of the problem and governing equations

The physical setup, illustrated in Fig. 1, comprises a differentially heated vertical porous layer saturated with an incompressible Newtonian fluid. Within this configuration, there exists a superimposed plane Couette flow component, induced by the motion of two vertical plates located at $x^* = \pm h$, which are moving with vertical velocities $\pm w_0$ and maintained at different constant temperatures $T_2$ and $T_1 (< T_2)$. The study adopts a Cartesian coordinate system, with the $x^*$-axis representing the horizontal direction perpendicular to the layer, the $y^*$-axis is also horizontal, and the $z^*$-axis being vertical. In this system, the starred symbols are used to represent the dimensional coordinates. The gravity is acting in the downward vertical direction. The flow in the porous medium is jointly driven by the movement of the vertical plates as well as due to the buoyancy force. The flow model is based on the local mass, momentum and energy balance equations according to Darcy's law with thermal equilibrium between the fluid and solid phases of the porous medium. Under the Oberbeck-Boussinesq approximation, the governing equations in the dimensionless form are

$$\nabla \cdot \vec{q} = 0, \qquad (1)$$

$$\frac{1}{Pr_D}\frac{\partial}{\partial t}(\nabla\times\vec{q}) = R_D\nabla\times(T\hat{k}) - \nabla\times\vec{q}, \qquad (2)$$

$$\frac{\partial T}{\partial t} + (\vec{q}\cdot\nabla)T = \nabla^2 T, \qquad (3)$$

The boundary conditions are

$$\vec{q}\cdot\hat{k} = \pm Pe \text{ at } x = \pm 1; \quad T = \pm 1/2 \text{ at } x = \pm 1. \qquad (4)$$

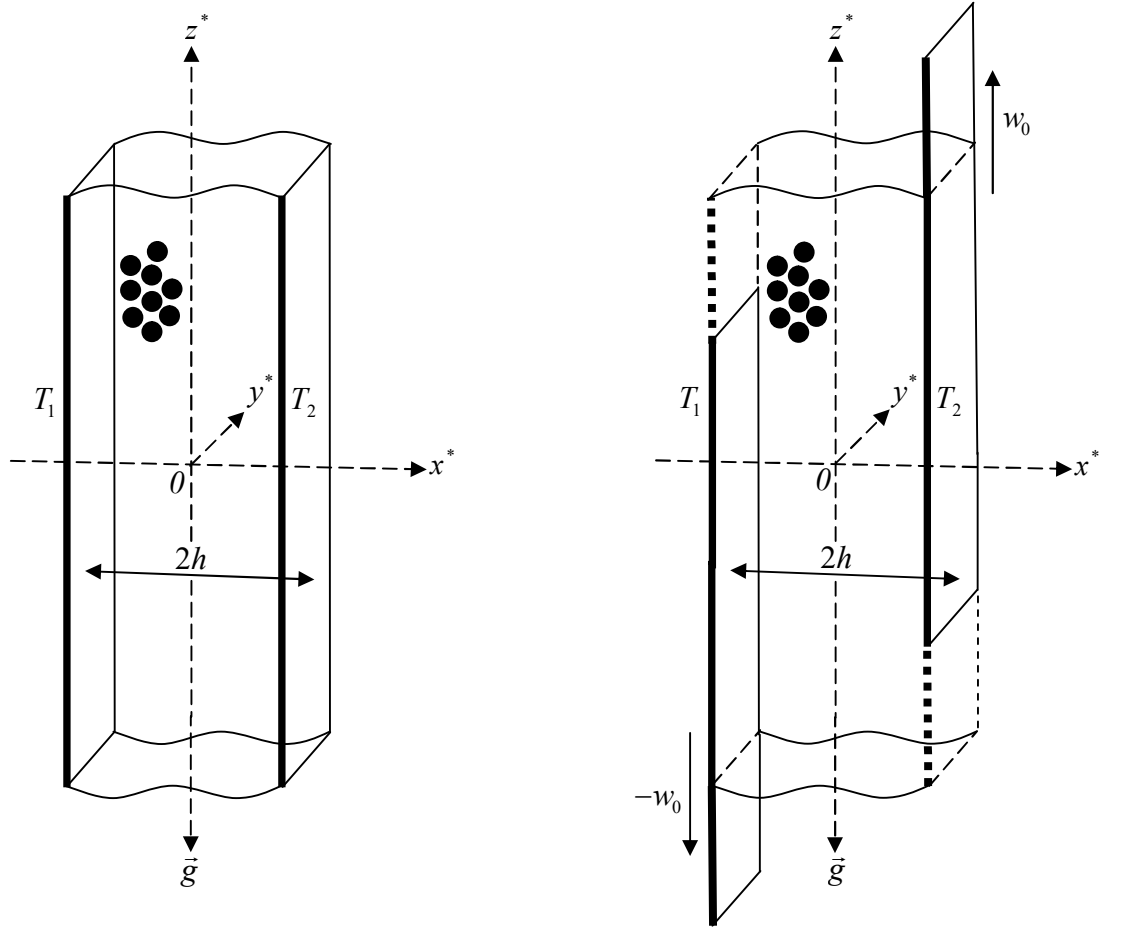

**Fig. 1.** Physical configuration of the problem.

The quantities in Eqs. (1)-(4) are the velocity, $\vec{q} = (u, v, w)$, temperature, $T$, time, $t$, the Darcy–Rayleigh number, $R_D = \rho_0 g \beta (T_2 - T_1) K h / \mu \kappa$, the Péclet number, $Pe = w_0 h / \kappa$, and the Prandtl-Darcy number, $Pr_D = \varepsilon \, Pr \, \sigma / Da$, while $\hat{k}$ is the unit vector in the vertical direction, $\rho_0$ is the density at the reference temperature $T_0 = (T_1 + T_2)/2$, $g$ is the modulus of the acceleration due to gravity, $\beta$ is the volumetric thermal expansion coefficient, $K$ is the permeability, $\mu$ is the fluid viscosity, $\kappa$ is the thermal diffusivity, $\varepsilon$ is the porosity, $\sigma$ is the ratio of volumetric heat capacity of the fluid-saturated porous medium to the volumetric heat capacity of the fluid, $Pr = \mu/\kappa\rho_0$ is the Prandtl number and $Da = K/h^2$ is the Darcy number. The non-dimensionalization was effected through the following transformations

$$(x^*, y^*, z^*)\frac{1}{h} = (x, y, z), \ (u^*, v^*, w^*)\frac{h}{\kappa} = (u, v, w), \ \frac{T^* - T_0}{T_2 - T_1} = T, \ \frac{\kappa}{\sigma h^2}t^* = t. \quad (5)$$

We consider two-dimensional motions and introduce the stream function $\psi(x, z, t)$ satisfying the continuity equation, such that

$$u = -\partial\psi/\partial z \text{ and } w = \partial\psi/\partial x. \quad (6)$$

Substituting Eq. (6) into Eqs. (2) and (3), we get

$$-\frac{1}{Pr_D}\frac{\partial}{\partial t}\left(\frac{\partial^2\psi}{\partial x^2} + \frac{\partial^2\psi}{\partial z^2}\right) = \left(\frac{\partial^2\psi}{\partial x^2} + \frac{\partial^2\psi}{\partial z^2}\right) - R_D\frac{\partial T}{\partial x}, \quad (7)$$

$$\frac{\partial T}{\partial t} + J(\psi, T) = \nabla^2 T, \quad (8)$$

where $J(\psi, T) = \frac{\partial\psi}{\partial x}\frac{\partial T}{\partial z} - \frac{\partial\psi}{\partial z}\frac{\partial T}{\partial x}$ is the Jacobian. (9)

The boundary conditions now become

$$\psi = \pm Pe \text{ at } x = \pm 1; \ T = \pm 1/2 \text{ at } x = \pm 1. \quad (10)$$

## 3. The basic stationary flow

A steady mixed convection flow is simplified by Eqs. (7), (8) and (10) when the velocity field is assumed to be fully developed. This leads to a basic solution given by

$$\psi_b(x) = \frac{R_D}{4}(x^2 - 1) + Pe\,x, \ T_b(x) = \frac{x}{2}. \quad (11)$$

where the subscript $b$ stands for the basic state. It is to be noted that first kind boundary conditions are being used in obtaining the basic state solution. The flow in the vertical porous layer is due to Couette flow if $R_D = 0$, and by buoyancy if $Pe = 0$ which coincides with that of Gill [1], Rees [5] and Barletta [8]. Figures 2 (a) and (b) illustrate the basic stream function $\psi_b(x)$ versus $x$ over the range $x \in [-1,1]$ for different values of $Pe$ when $R_D = 10$ and 100, respectively. A focus on the case of $Pe \neq 0$ has been made as there is a major qualitative difference, shown in Fig. 2(a), between the plates being at rest and in relative motion. It is observed that as $Pe$ increases, the basic stream function profiles gradually turn out to be linear from parabolic (Fig. 2a). It is also noticed that the maximum value of $\psi_b$ increases with increasing $Pe$, which adheres to the right wall and a delay in this process is evident from Fig. 2(b) as $R_D$ increases.

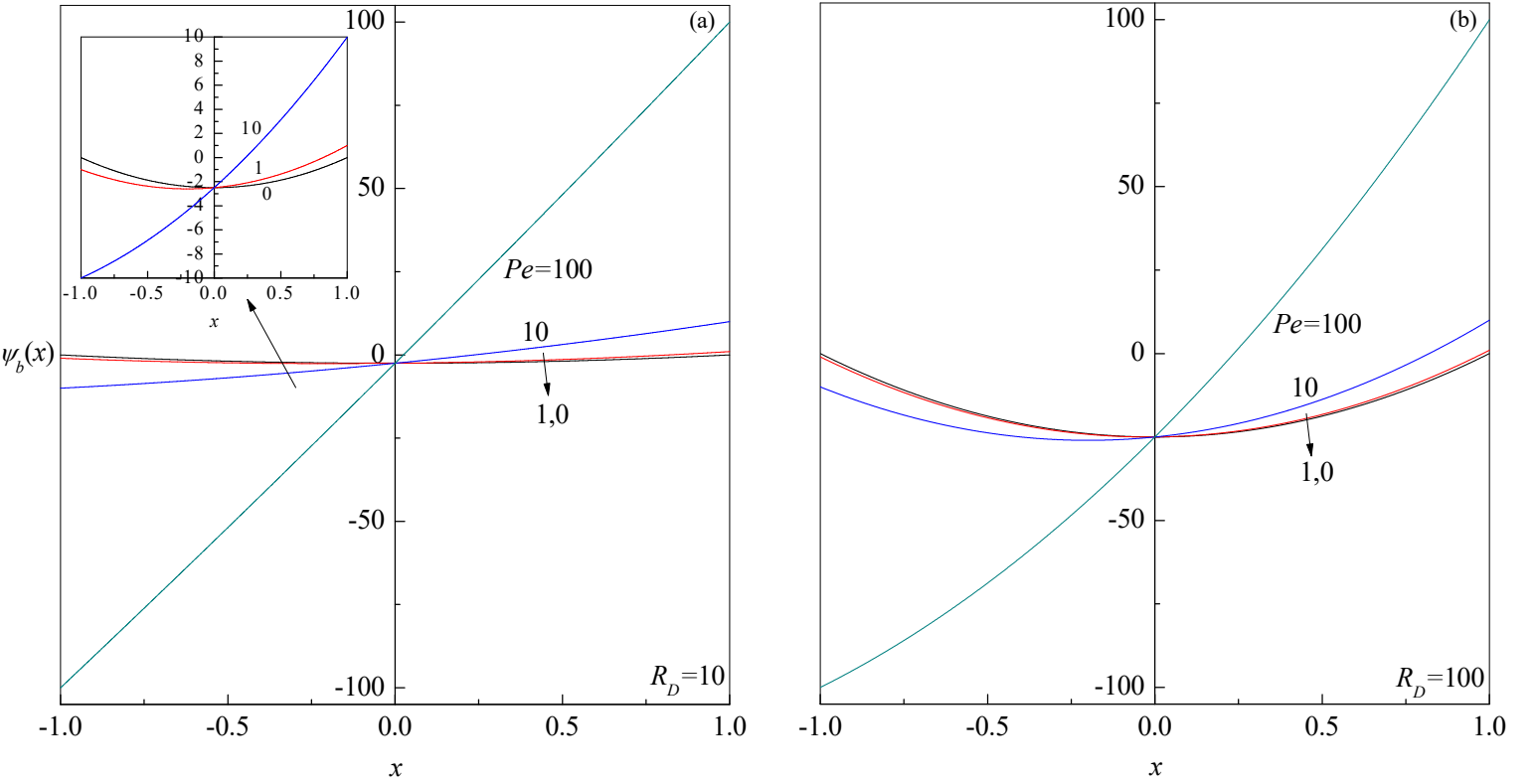

**Fig. 2.** Plots of the basic stream function for different values of $Pe$ when (a) $R_D = 10$ and (b) $R_D = 100$.

## 4. Linear stability analysis

To perform linear stability analysis, we proceed to perturb the basic state by applying small-amplitude disturbances, which are periodic in the vertical direction

$$(\psi, T)(x, z, t) = (\psi_b(x), T_b(x)) + (\Psi, \Theta)(x) e^{ia(z-ct)}, \tag{12}$$

with real wave number $a$ and the complex wave speed $c (= c_r + ic_i)$. Substituting Eq. (12) into Eqs. (7) and (8) and using Eq. (11), the disturbance quantities $\Psi(x)$ and $\Theta(x)$ are found to satisfy the following stability equations

$$\frac{iac}{Pr_D}(D^2 - a^2)\Psi = (D^2 - a^2)\Psi - R_D D\Theta, \tag{13}$$

$$(D^2 - a^2)\Theta + \frac{ia}{2}\Psi - iaD\psi_b \Theta = -iac\Theta, \tag{14}$$

where $D = d/dx$ is the differential operator.

Since the nature of velocity and temperature boundary conditions alter significantly the stability characteristics of fluid flow, we impose Robin boundary conditions for $\Psi(x)$ and $\Theta(x)$ in the form

$$N_0 D\Psi \pm \Psi = 0;\ Bi\, D\Theta \pm \Theta = 0\ \text{at}\ x = \pm 1, \tag{15}$$

where $N_0$ is a non-negative parameter and $Bi$ is the Biot number. The limiting case $N_0 = 0$ and $N_0 \to \infty$ correspond to impermeable and permeable boundary conditions, respectively while the intermediate values of $N_0$ relate to the partial permeable condition. The parameter $Bi$ modulates the imperfectly conducting boundaries including the limiting cases $Bi = 0$ (isothermal) and $Bi \to \infty$ (adiabatic).

## 5. Numerical solution

The basis for developing the stability analysis is the eigenvalue problem constituted by Eqs. (13)-(15), which has been solved by employing the Chebyshev collocation method. Accordingly, $\Psi$ and $\Theta$ are expanded as

$$\Psi = \sum_{n=0}^{N-1} a_n \Psi_n(x),\ \Theta = \sum_{n=0}^{N-1} b_n \Theta_n(x), \tag{16}$$

where $a_n$ and $b_n$ are constants, while $\Psi_n(x)$ and $\Theta_n(x)$ are chosen satisfying the respective boundary conditions in the form

$$\Psi_n(x) = \left(1 - x^2 + \frac{2N_0}{1 + N_0 n^2}\right) \tau_n(x),\ \Theta_n(x) = \left(1 - x^2 + \frac{2Bi}{1 + Bi\, n^2}\right) \tau_n(x). \tag{17}$$

Here $\tau_n(x)$ is the Chebyshev polynomial of order $n$ defined by

$$\tau_0(x) = 1;\ \tau_1(x) = x;\ \tau_{n+1}(x) - 2x\tau_n(x) + \tau_{n-1}(x) = 0\ (n \geq 1). \tag{18}$$

Substituting Eq. (17) back into Eqs. (13) and (14) and requiring that they are satisfied at $N$ collocation points $x_0, x_1, x_2, \ldots, x_{N-1}$, where

$$x_m = \cos\{[1 - (m+1)/(N+1)]\pi\},\ m = 0, 1, 2, \cdots, N-1, \tag{19}$$

leads to a system of algebraic equations which can be written in the matrix form

$$AX = cBX. \tag{20}$$

Here, $X^T = (a_0, a_1, a_2, \ldots, a_{N-1}, b_0, b_1, b_2, \ldots, b_{N-1})$ is the transpose of the column vector $X$ and the coefficient matrices $A$ and $B$ are of dimension $2N \times 2N$ given by

$$A = \begin{pmatrix} A_{11} & A_{12} \\ A_{21} & A_{22} \end{pmatrix};\ B = \begin{pmatrix} B_{11} & B_{12} \\ B_{21} & B_{22} \end{pmatrix}, \tag{21}$$

where

$$A_{11}(m,n) = (D^2 - a^2)\Psi_n(x_m),$$

$$A_{12}(m,n) = -R_D D\Theta_n(x_m),$$

$$A_{21}(m,n) = -ia\, DT_b(x_m)\Psi_n(x_m),$$

$$A_{22}(m,n) = ia\, D\psi_b(x_m)\Theta_n(x_m) - (D^2 - a^2)\Theta_n(x_m),$$

$$B_{11}(m,n) = \frac{ia}{Pr_D}(D^2 - a^2)\Psi_n(x_m),$$

$$B_{12}(m,n) = B_{21}(m,n) = 0,$$

$$B_{22}(m,n) = ia\,\Theta_n(x_m). \tag{22}$$

For fixed values of $R_D, Pr_D, Pe$ and $a$, the values of $c$ which ensure a non-trivial solution of Eq. (20) are obtained as the eigenvalues of the matrix $B^{-1}A$.

## 6. Discussion of the results

The stability characteristics of the base flow are examined separately for different types of velocity and temperature boundary conditions through the parameters $N_0$ and $Bi$, respectively. The other factors influencing the instability of the base flow are the Darcy-Rayleigh number $R_D$, the Prandtl-Darcy number $Pr_D$ and the Péclet number $Pe$. The eigenvalue problem is solved numerically by employing the Chebyshev collocation method to analyze the stability/instability of the base flow.

### 6.1 Boundaries are impermeable and isothermal $(N_0 = 0 = Bi)$

The most unstable modes computed for various combinations of $Pe, Pr_D, R_D$ and $a$ by varying the collocation points $N$ are presented in Table 1. Notably, the table highlights a rapid convergence of the eigenvalue as $N$ increases. The temporal growth rate $c_i$ of the perturbations is computed as a function of wave number $a$ for a wide range of governing parameters and the results are displayed in Figs. 3(a-c) for some representative values of $Pe, R_D$ and $Pr_D$. The results were calculated using $N = 70$. From the figures, it is evidenced that $c_i$ remains negative for all values of $a$ indicating that no instability is possible, which is akin to the result perceived by Rees [4] for the fixed plates case (i.e., $Pe = 0$). The selection of less/more stable modes largely depends on the value of $a$ and in fact $c_i$ displays a weak dependence at higher values of $a$ for different values of physical parameters. Moreover, it is obvious that $Pe$ has insignificant influence on $c_i$ irrespective of the values of $a$.

| N | $Pr_D=1, Pe=1, a=1, R_D=10^3$ | | $Pr_D=1, Pe=10^2, a=1, R_D=10^3$ | | $Pr_D=1, Pe=1, a=1, R_D=10^5$ | | $Pr_D=10^3, Pe=1, a=1, R_D=10^3$ | | $Pr_D=1, Pe=1, a=2, R_D=10^3$ | |
|---|---|---|---|---|---|---|---|---|---|---|
|  | $c_i$ | $c_r$ | $c_i$ | $c_r$ | $c_i$ | $c_r$ | $c_i$ | $c_r$ | $c_i$ | $c_r$ |
| 5 | -0.723174 | -0.051145 | -0.713294 | -0.007542 | -0.722145 | -0.053461 | -15.469866 | 1.000834 | -0.386684 | -0.035945 |
| 10 | -0.734328 | -0.000524 | -0.716650 | 0.016106 | -0.721715 | 0 | -44.609149 | -73.316887 | -0.396456 | -0.027676 |
| 15 | -0.721376 | 0.000163 | -0.717484 | 0.019341 | -0.708227 | 0.000150 | -62.527848 | 407.023579 | -0.401809 | -0.024619 |
| 20 | -0.721333 | 0.000162 | -0.717527 | 0.019048 | -0.721428 | 0 | -62.871111 | 406.779498 | -0.406804 | -0.029484 |
| 25 | -0.721334 | 0.000162 | -0.717528 | 0.019049 | -0.737750 | -0.010257 | -62.873914 | 406.781576 | -0.406757 | -0.029698 |
| 30 | -0.721334 | 0.000162 | -0.717528 | 0.019049 | -0.722993 | 0 | -62.873947 | 406.781584 | -0.406757 | -0.029698 |
| 35 |  |  |  |  | -0.722338 | 0 | -62.873947 | 406.781584 |  |  |
| 40 |  |  |  |  | -0.721180 | 0 |  |  |  |  |
| 45 |  |  |  |  | -0.720909 | 0 |  |  |  |  |
| 50 |  |  |  |  | -0.721698 | 0 |  |  |  |  |
| 55 |  |  |  |  | -0.721723 | 0 |  |  |  |  |
| 60 |  |  |  |  | -0.721701 | 0 |  |  |  |  |
| 65 |  |  |  |  | -0.721701 | 0 |  |  |  |  |

**Table 1.** Chebyshev approximation to the most unstable mode for different set of parameters when $N_0=0$ and $Bi=0$.

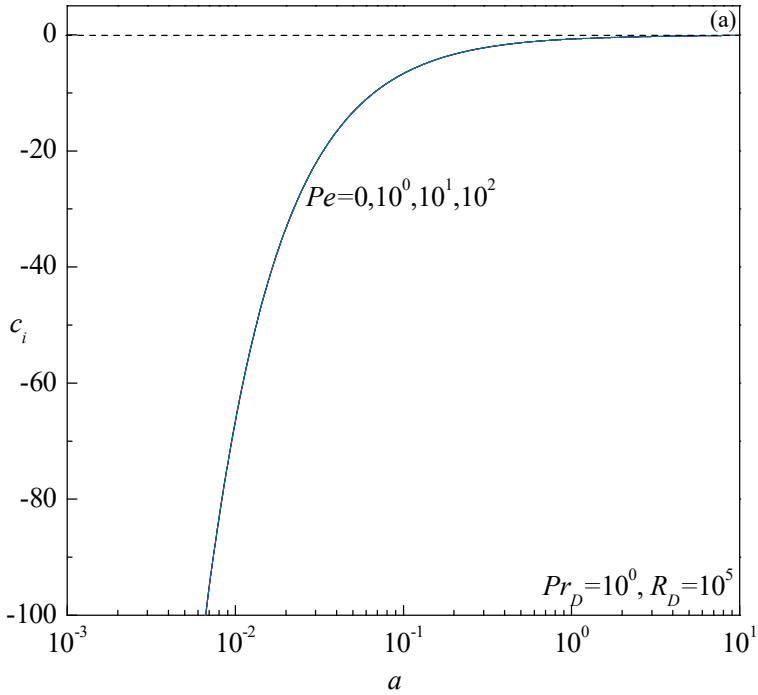
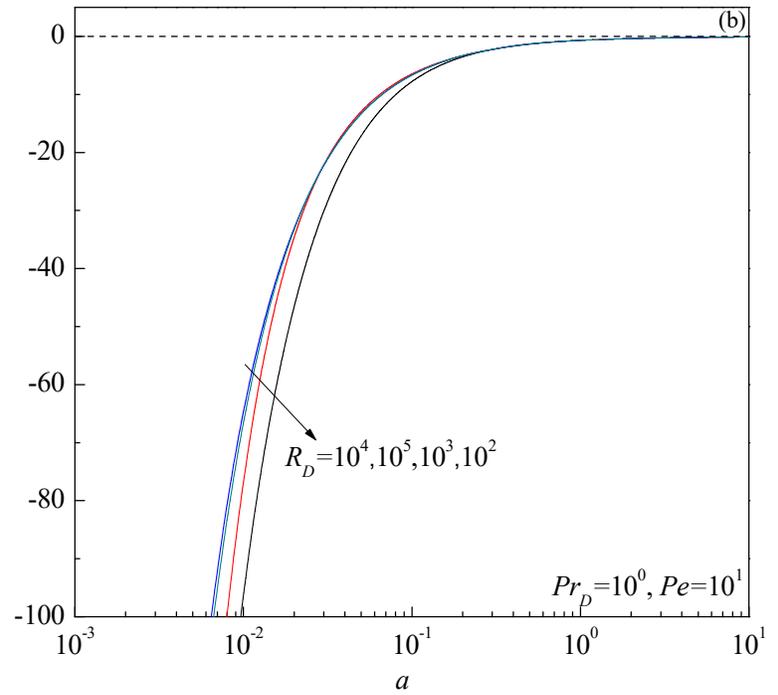

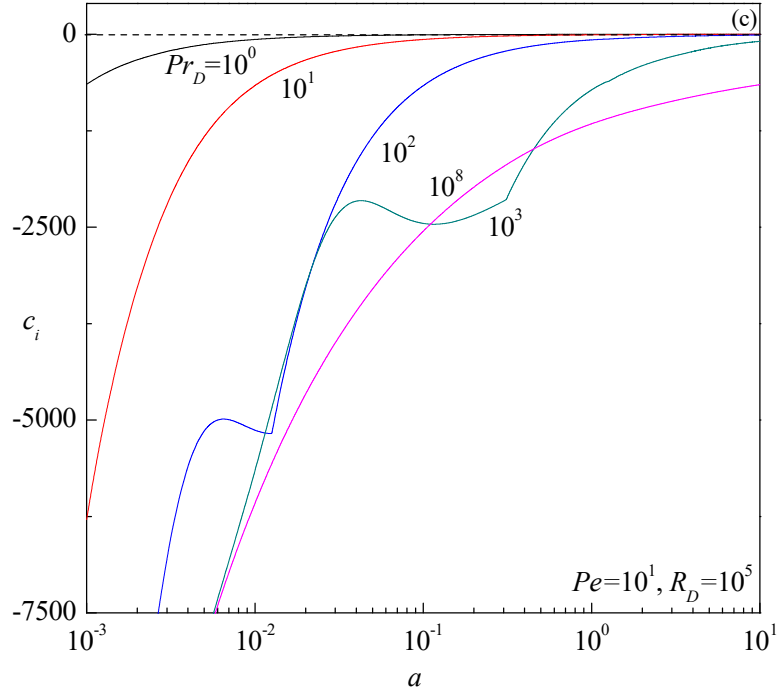

**Fig. 3.** Plots of growth rate $c_i$ versus $a$ for different values of (a) $Pe$, (b) $R_D$ and (c) $Pr_D$.

As observed from Fig. 3(c), the base flow is stable in the limit $Pr_D \to \infty$ and this has prompted us to examine the stability of base flow analytically. To proceed with, the integral method of Gill [1] is employed to find the sign of the growth rate $c_i$ without the time-dependent velocity term in the momentum equation. Accordingly, we operate $(D^2 - a^2)$ on Eq. (14) and make use of Eq. (13) so that we can write a single fourth-order differential equation for $\Theta$ in the form

$$\left(\Theta'''' + a^4\Theta - 2a^2\Theta''\right) - \frac{iaR_D}{2}\left((x\Theta')' - a^2 x\Theta\right) - iaPe\left(\Theta'' - a^2\Theta\right) = -iac\left(\Theta'' - a^2\Theta\right). \quad (23)$$

Multiplying Eq. (23) by $\bar{\Theta}$, the complex conjugate of $\Theta$, and integrating it over $x \in [-1,1]$, we then obtain the following equation

$$\int_{-1}^{1}\left(\Theta''''\bar{\Theta} + a^4|\Theta|^2 - 2a^2\Theta''\bar{\Theta}\right)dx - \frac{iaR_D}{2}\int_{-1}^{1}\left[(x\Theta')'\bar{\Theta} - a^2 x|\Theta|^2\right]dx - iaPe\int_{-1}^{1}\left(\Theta''\bar{\Theta} - a^2|\Theta|^2\right)dx$$
$$= -iac\int_{-1}^{1}\left(\Theta''\bar{\Theta} - a^2|\Theta|^2\right)dx. \quad (24)$$

Integrating Eq. (24) by parts and using the boundary conditions $\Theta(\pm 1) = 0$, we get

$$\int_{-1}^{1}\left(|\Theta''|^2+a^4|\Theta|^2+2a^2|\Theta'|^2\right)dx+\frac{iaR_D}{2}\int_{-1}^{1}x\left(|\Theta'|^2+a^2|\Theta|^2\right)dx+iaPe\int_{-1}^{1}\left(|\Theta'|^2+a^2|\Theta|^2\right)dx$$
$$=iac\int_{-1}^{1}\left(|\Theta'|^2+a^2|\Theta|^2\right)dx. \tag{25}$$

Equating the real part of Eq. (25) leads to the following relation

$$\int_{-1}^{1}\left(|\Theta''|^2+a^4|\Theta|^2+2a^2|\Theta'|^2\right)dx=-ac_i\int_{-1}^{1}\left(|\Theta'|^2+a^2|\Theta|^2\right)dx. \tag{26}$$

Equation (26) is exactly the same as the one obtained by Gill [1], which allows one to conclude that $c_i$ is always strictly negative for all infinitesimal perturbations. Thus, we reach at the conclusion that the base flow is asymptotically stable even if the plates are in relative motion when the boundaries are impermeable-isothermal.

### 6.2 Boundaries are impermeable and imperfectly conducting ($N_0 = 0, Bi \neq 0$ and $Bi \to \infty$)

Here, Robin type of temperature boundary conditions are considered so as to analyse accurately the initiation of instability, if at all exists, with the change in the Biot number along with the other physical parameters. The present set up describes heat transfer to the external environment, with a finite conductance.

### 6.2.1 Growth rate

The results of the most unstable mode for different combinations of governing parameters are given in Table 2, aiming to assist others who may employ this calculation as a reference point. It is observed that accurate results can be achieved with lower values of $N$ as $R_D$ decreases. The evaluated growth rate $c_i$ of normal mode perturbations are displayed in Figs. 4(a-d) as a function of $a$ for different values of $Pe, R_D, Bi$ and $Pr_D$ by taking $N = 50$. These figures show the possibility of flow becoming unstable as the value of $c_i$ changes from negative to positive in some cases. It is intriguing to note from Figs. 4(a, c) that the instability occurs whether the plates are moving ($Pe \neq 0$) or not ($Pe = 0$) when the boundaries are either imperfectly conducting $(Bi \neq 0)$ or perfectly adiabatic ($Bi \to \infty$). These findings stand in stark contrast to the situation discussed in the previous subsection (Sec. 6.1) regarding impermeable-isothermal boundaries. Consequently, it represents one of the most captivating breakthroughs of Gill's problem.

| N | $R_D = 10^3, a=1, Pe=1,$ $Bi=5, Pr_D=800$ | | $R_D = 10^3, a=1, Pe=10^2,$ $Bi=5, Pr_D=800$ | | $R_D = 10^4, a=1, Pe=1,$ $Bi=5, Pr_D=800$ | | $R_D = 10^3, a=1, Pe=1,$ $Bi=20, Pr_D=800$ | | $R_D = 10^3, a=1, Pe=1,$ $Bi=5, Pr_D=10^4$ | |
|---|---|---|---|---|---|---|---|---|---|---|
| | $c_i$ | $c_r$ | $c_i$ | $c_r$ | $c_i$ | $c_r$ | $c_i$ | $c_r$ | $c_i$ | $c_r$ |
| 5 | 5.402073 | 467.980099 | 7.249044 | 564.622018 | 86.108880 | -4353.966597 | 6.161489 | 468.258696 | -7.150419 | 474.722035 |
| 10 | -8.737390 | 471.395490 | -7.865134 | 567.750081 | -12.785775 | -2105.030680 | -7.806344 | 471.846355 | -17.585372 | 481.178215 |
| 15 | -8.077647 | 471.295878 | -7.091324 | 567.699479 | -75.773957 | -1936.592035 | -7.157253 | 471.755215 | -17.337558 | 480.807108 |
| 20 | -8.050150 | 471.219552 | -7.052605 | 567.614510 | -98.140417 | -4725.553371 | -7.131984 | 471.680771 | -17.361027 | 480.757602 |
| 25 | -8.052975 | 471.216895 | -7.055566 | 567.611067 | -106.543201 | -4726.409976 | -7.134746 | 471.678264 | -17.363606 | 480.758196 |
| 30 | -8.053073 | 471.216876 | -7.055679 | 567.611031 | -104.856469 | -4726.403350 | -7.134840 | 471.678247 | -17.363642 | 480.758251 |
| 35 | -8.053075 | 471.216875 | -7.055681 | 567.611030 | -105.111436 | -4726.206810 | -7.134841 | 471.678247 | -17.363642 | 480.758252 |
| 40 | -8.053075 | 471.216875 | -7.055681 | 567.611030 | -105.108407 | -4726.256005 | -7.134841 | 471.678247 | | |
| 45 | | | | | -105.102967 | -4726.253028 | | | | |
| 50 | | | | | -105.103382 | -4726.252532 | | | | |
| 55 | | | | | -105.103426 | -4726.252566 | | | | |
| 60 | | | | | -105.103424 | -4726.252570 | | | | |
| 65 | | | | | -105.103424 | -4726.252570 | | | | |

**Table 2.** Chebyshev approximation to the most unstable mode for different set of parameters when $N_0 = 0$.

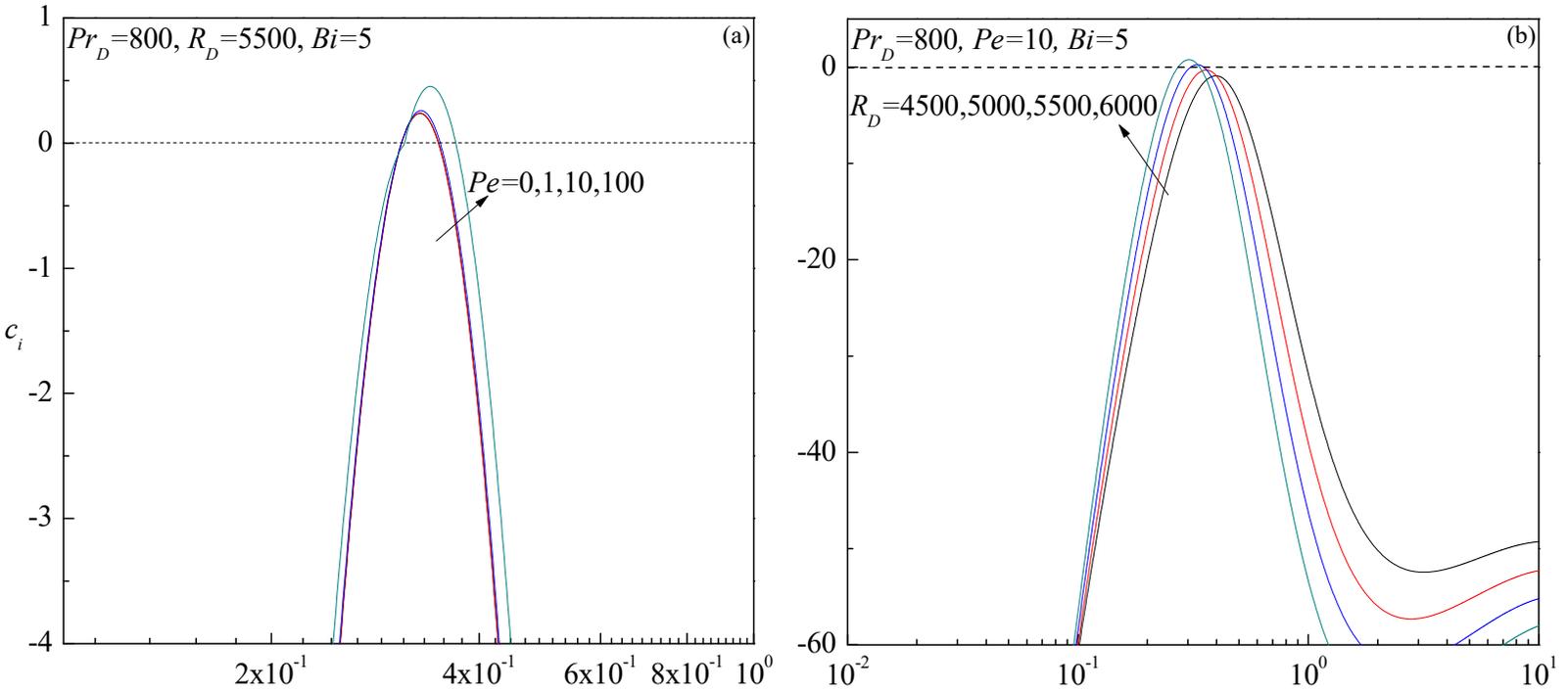

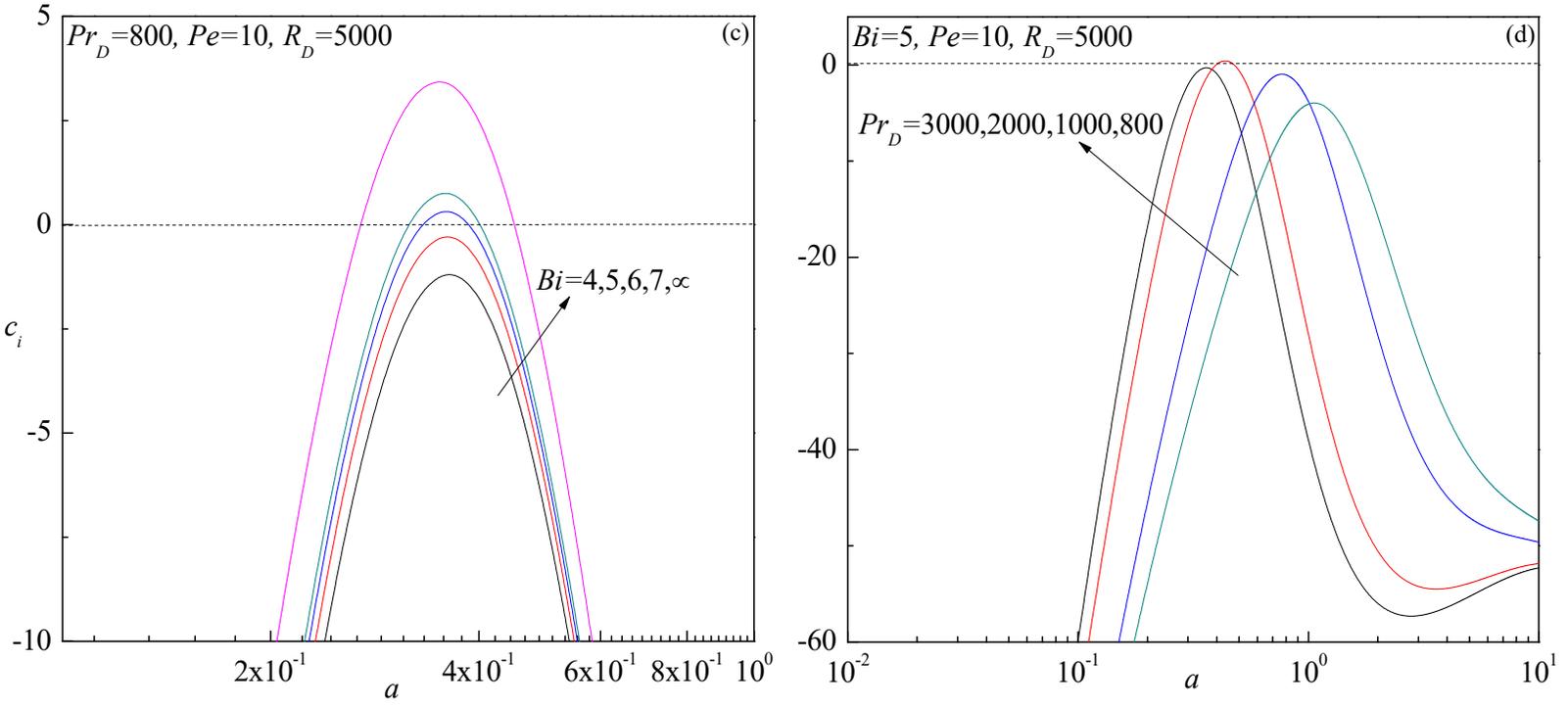

**Fig. 4.** Plots of growth rate $c_i$ versus $a$ for different values of (a) $Pe$, (b) $R_D$, (c) $Bi$ and (d) $Pr_D$.

### 6.2.2 Neutral stability curves

The threshold value of $R_D$ for the onset of modal or convective instability is determined by examining the long-time behavior of each single Fourier mode. The neutral stability curves represented in the $(a, R_D)$-plane depict the state of zero growth rate. At the nadir of the neutral stability curve within the $(a, R_D)$-plane lies the critical parameter combination that marks the onset of instability. Hence, the minimum value of $R_D$ along each curve defines the critical value $(a_c, R_{Dc})$ which is influenced by $Pe, Pr_D$ and $Bi$. Instability is possible when $R_D$ exceeds $R_{Dc}$. In order to achieve these results, Eqs. (13)-(15) are solved numerically. From $2N$ eigenvalues one having the largest imaginary part, say $c_1 = c_{r1} + ic_{i1}$ is selected. We then enforce $c_{i1} = 0$ by varying either $R_D$ with $a$ fixed, or $a$ with $R_D$ fixed to obtain the neutral stability point. The infimum of $R_D$ as a function of $a$ or vice versa gives the critical Darcy-Rayleigh number $R_{Dc}$ and the critical wave number $a_c$. The real part of $c_1$ i.e., $c_{r1}$ gives the critical wave speed denoted by $c_c$. If zero imaginary part of $c_1$ are characterized by a zero real part of $c_1$ then the phase velocity of the neutrally stable modes is zero. This means that the transition to instability occurs with stationary modes. If real part of $c_1$ is non-zero then the modes are travelling-wave.

The critical stability parameters are tabulated in Table 3 for different sets of governing parameters by varying the collocation points $N$ to check the performance and the accuracy of the numerical method employed. From Table 3, it is apparent that lesser number of collocation points are sufficient to achieve the desired degree of accuracy as $Pr_D$ decreases. The numerical outcomes provided in this study have a precision of up to six decimal places.

| N | $Pr_D = 10^3, Pe = 10, Bi = 20$ | | | $Pr_D = 10^3, Pe = 10, Bi = 3$ | | | $Pr_D = 10^3, Pe = 10^2, Bi = 20$ | | | $Pr_D = 10^4, Pe = 10, Bi = 20$ | | |
|---|---|---|---|---|---|---|---|---|---|---|---|---|
| | $R_{Dc}$ | $a_c$ | $c_c$ | $R_{Dc}$ | $a_c$ | $c_c$ | $R_{Dc}$ | $a_c$ | $c_c$ | $R_{Dc}$ | $a_c$ | $c_c$ |
| 5 | 499.986646 | 2.915221 | 237.430863 | 530.574783 | 2.938518 | 250.721465 | 447.372549 | 2.649513 | 303.164881 | 1346.792124 | 4.089162 | 623.564272 |
| 10 | 4883.183013 | 0.414485 | 2313.803302 | 13416.233136 | 0.877435 | -2823.372444 | 4848.176251 | 0.406920 | 2385.229478 | 3881.667902 | 6.091444 | 1893.556407 |
| 15 | 3329.241066 | 0.633208 | 1579.504384 | 4953.814000 | 0.465673 | -2320.921382 | 3319.893255 | 0.606875 | 1663.254747 | 12817.579265 | 1.268885 | 6258.820184 |
| 20 | 3704.958223 | 0.552125 | 1757.376194 | 6535.836523 | 0.343362 | -3066.523805 | 3668.776764 | 0.534656 | 1828.382870 | 11178.133497 | 1.437809 | 5456.748608 |
| 25 | 3710.434832 | 0.553061 | 1759.958303 | 6520.532455 | 0.345684 | -3059.291098 | 3679.146717 | 0.534589 | 1833.284027 | 12105.888774 | 1.366120 | 5911.056860 |
| 30 | 3708.240878 | 0.553564 | 1758.914783 | 6506.219475 | 0.346555 | -3052.536111 | 3677.324207 | 0.535006 | 1832.416963 | 11901.745116 | 1.386153 | 5810.967108 |
| 35 | 3708.155119 | 0.553582 | 1758.873863 | 6505.819398 | 0.346572 | -3052.347724 | 3677.241109 | 0.535022 | 1832.377425 | 11949.555463 | 1.381602 | 5834.384301 |
| 40 | 3708.153794 | 0.553580 | 1758.873354 | 6505.819570 | 0.346572 | -3052.347781 | 3677.239494 | 0.535022 | 1832.376660 | 11938.809456 | 1.383584 | 5829.102685 |
| 45 | 3708.153777 | 0.553580 | 1758.873342 | 6505.819712 | 0.346572 | -3052.347856 | 3677.239468 | 0.535026 | 1832.376471 | 11940.464034 | 1.383032 | 5829.921889 |
| 50 | 3708.153778 | 0.553580 | 1758.873347 | 6505.819717 | 0.346572 | -3052.347817 | 3677.239468 | 0.535022 | 1832.376662 | 11940.375925 | 1.383131 | 5829.876271 |
| 55 | 3708.153778 | 0.553580 | 1758.873367 | 6505.819715 | 0.346572 | -3052.347820 | 3677.239468 | 0.535022 | 1832.376677 | 11940.359234 | 1.383124 | 5829.868362 |
| 60 | 3708.153778 | 0.553579 | 1758.873430 | 6505.819716 | 0.346572 | -3052.347869 | 3677.239468 | 0.535022 | 1832.376663 | 11940.361698 | 1.383119 | 5829.869739 |
| 65 | 3708.153774 | 0.553579 | 1758.873431 | 6505.819716 | 0.346572 | -3052.347822 | 3677.239468 | 0.535022 | 1832.376644 | 11940.361749 | 1.383126 | 5829.869528 |
| 70 | 3708.153778 | 0.553580 | 1758.873362 | 6505.819716 | 0.346572 | -3052.347821 | 3677.239468 | 0.535022 | 1832.376679 | 11940.361738 | 1.383126 | 5829.869522 |
| 75 | 3708.153778 | 0.553580 | 1758.873375 | 6505.819716 | 0.346572 | -3052.347839 | 3677.239468 | 0.535023 | 1832.376635 | 11940.361739 | 1.383120 | 5829.869715 |
| 80 | 3708.153778 | 0.553580 | 1758.873370 | 6505.819716 | 0.346572 | -3052.347842 | 3677.239468 | 0.535023 | 1832.376603 | 11940.361735 | 1.383125 | 5829.869547 |
| 85 | 3708.153778 | 0.553580 | 1758.873377 | 6505.819715 | 0.346572 | -3052.347960 | 3677.239468 | 0.535023 | 1832.376625 | 11940.361734 | 1.383124 | 5829.869588 |
| 90 | 3708.153778 | 0.553580 | 1758.873356 | 6505.819714 | 0.346572 | -3052.347907 | 3677.239468 | 0.535023 | 1832.376611 | 11940.361737 | 1.383127 | 5829.869477 |
| 95 | 3708.153778 | 0.553579 | 1758.873402 | | | | | | | 11940.361718 | 1.383127 | 5829.869473 |
| 100 | 3708.153777 | 0.553580 | 1758.873382 | | | | | | | 11940.361722 | 1.383128 | 5829.869457 |
| 105 | | | | | | | | | | 11940.361732 | 1.383120 | 5829.869722 |
| 110 | | | | | | | | | | 11940.361731 | 1.383130 | 5829.869360 |
| 115 | | | | | | | | | | 11940.361717 | 1.383128 | 5829.869435 |
| 120 | | | | | | | | | | 11940.361734 | 1.383127 | 5829.869482 |

**Table 3.** Process of convergence of the Chebyshev collocation method when $N_0 = 0$.

Figures 5(a-d) display some neutral stability curves for different values of $Pe, Pr_D$ and $Bi$. The instability region gets enlarged with increasing $Pe$ (Fig. 5a) and $Bi$ (Fig. 5c). This is accompanied by a downward shift in the neutral stability curves and the critical point drifts leftward (rightward) with increasing $Pe$ ($Bi$). Interestingly, the neutral curves exhibit a bi-modal shape but only at higher values of $Pe$ and the preferred mode of instability, which corresponds to the global minimum with $c_c > 0$, is found in the lower wave number region (Fig. 5a). Moreover, as $Bi$ decreases, the width of the instability region strip gradually becomes narrower, eventually approaching the line $a = 0$ (Fig. 5c). As $Pr_D$ becomes larger, the onset of instability necessitates both smaller and larger Darcy-Rayleigh number associated with smaller and smaller wavelengths

(Fig. 5b). The neutral curves exemplified in Fig. 5(d) for different values of $Bi$ consist of two different branches connecting together in which the first and second ones correspond to positive and negative values of $c$, respectively and each of them attains their own minimum. For $Bi = 6.5$ and 7, the mode characterized by $c_c < 0$ has a lower value of $R_{Dc}$ compared to the mode with $c_c > 0$ suggesting that the unstable mode propagates in a downward direction. Nonetheless, for $Bi = 7.5$ and 8 the trend gets reversed as the dominant mode of instability shifts to the lower wave number region.

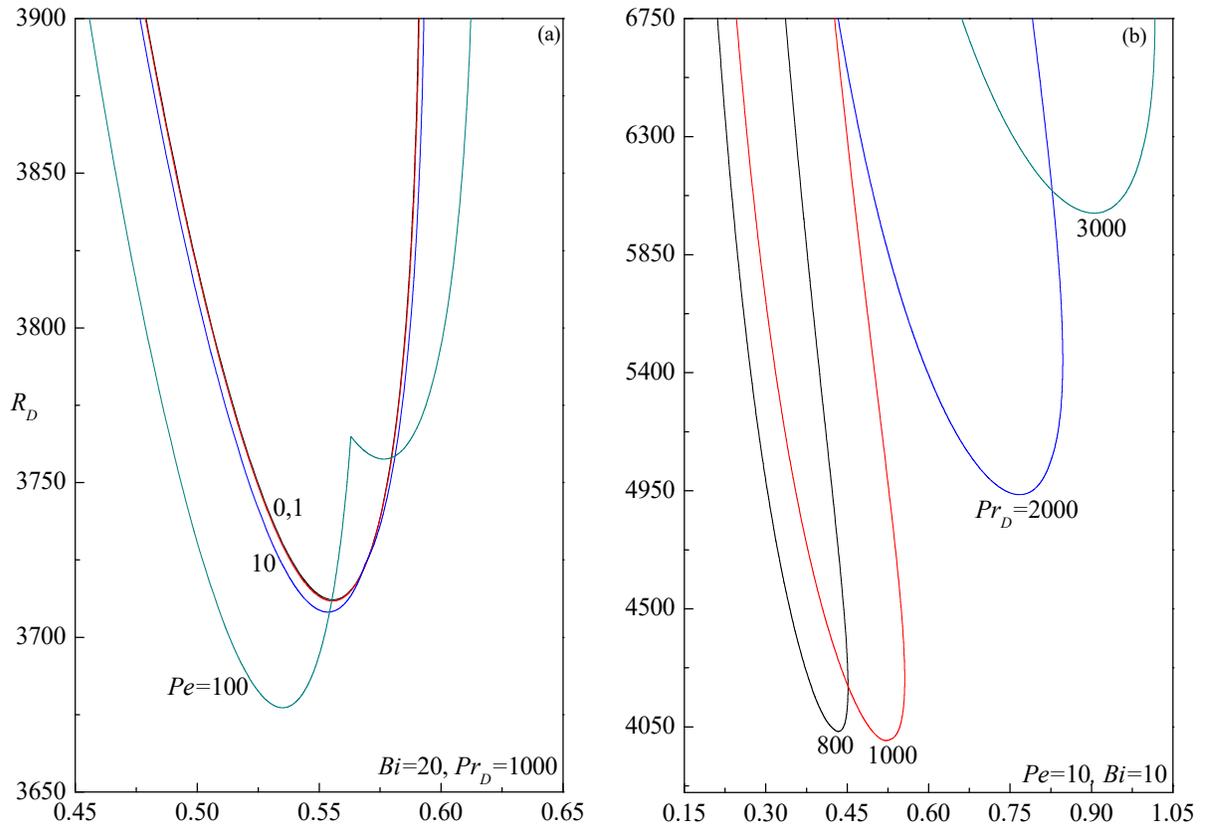

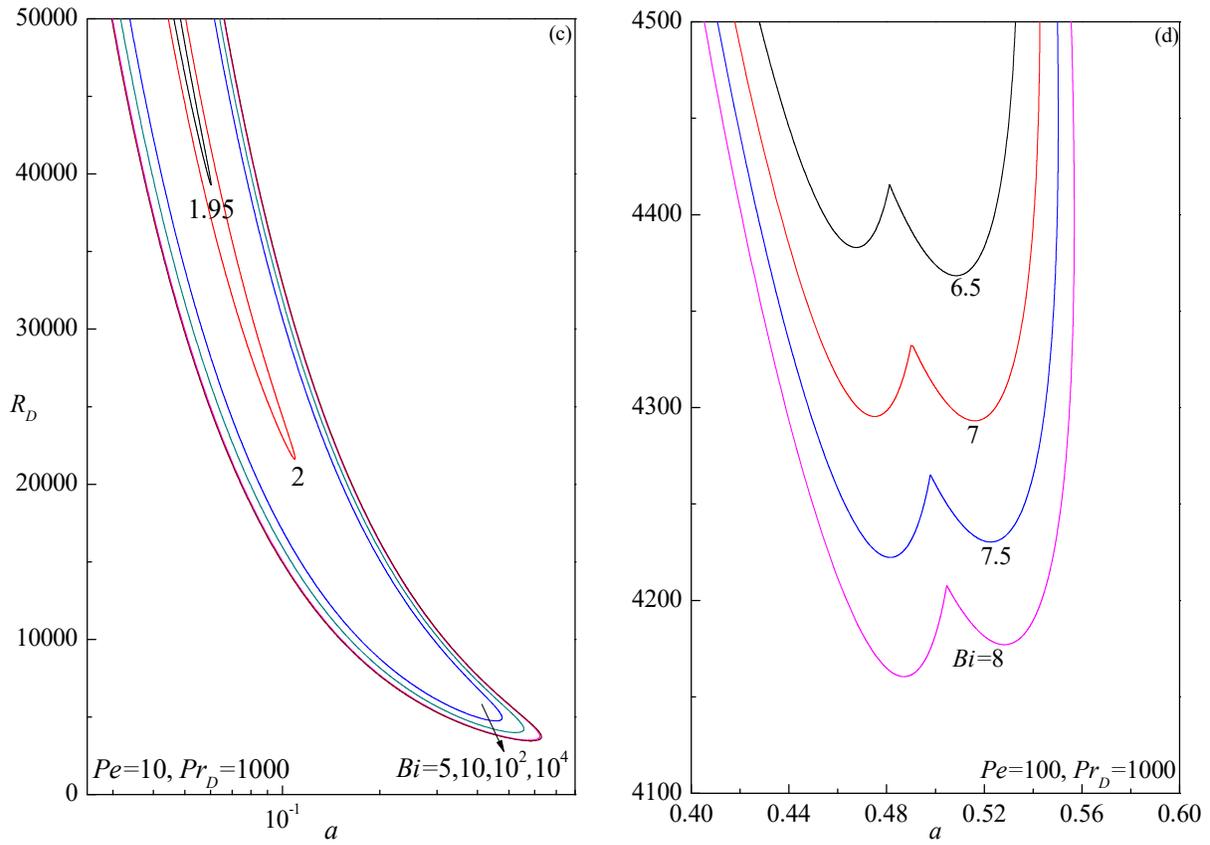

**Fig. 5.** Neutral stability curves in the $(a, R_D)$-plane for different values of (a) $Pe$, (b) $Pr_D$ and (c, d) $Bi$.

### 6.2.3 Condition for instability

The trend of critical triplets $R_{Dc}, a_c$ and $c_c$ with $Bi$ for different values of $Pr_D$ is presented graphically in Figs. 6, 7 and 8, respectively and each of these figures demonstrates the outcomes for four different values of $Pe = 0$, 1, 10 and 100. From Fig. 6(a), it is evident that there exists a threshold value of $Bi$, for each value of $Pr_D$, prior to which the base flow remains stable as $R_{Dc} \to \infty$ and the sign of growth rate is negative. Beyond this threshold value, however, a self-excited mode of disturbance emerges for both partially isothermal and perfectly adiabatic conditions. The value of $R_{Dc}$ decreases rapidly as $Bi$ increases and thereafter it remains unaffected with further increase in $Bi$. Thus, the adiabatic boundaries activate instability much faster than partial conducting boundaries. The effect of $Pr_D$ is both stabilizing and destabilizing on the base flow depending on the values of $Bi$. A similar kind of bearing could be seen for $Pe \neq 0$ (Figs. 6b-d) and note that it is challenging to differentiate the results when $Pe$ takes on values of both 0 and 1. However, $R_{Dc}$ decreases at higher values of $Pe$ indicating its effect is to hasten the onset of instability.

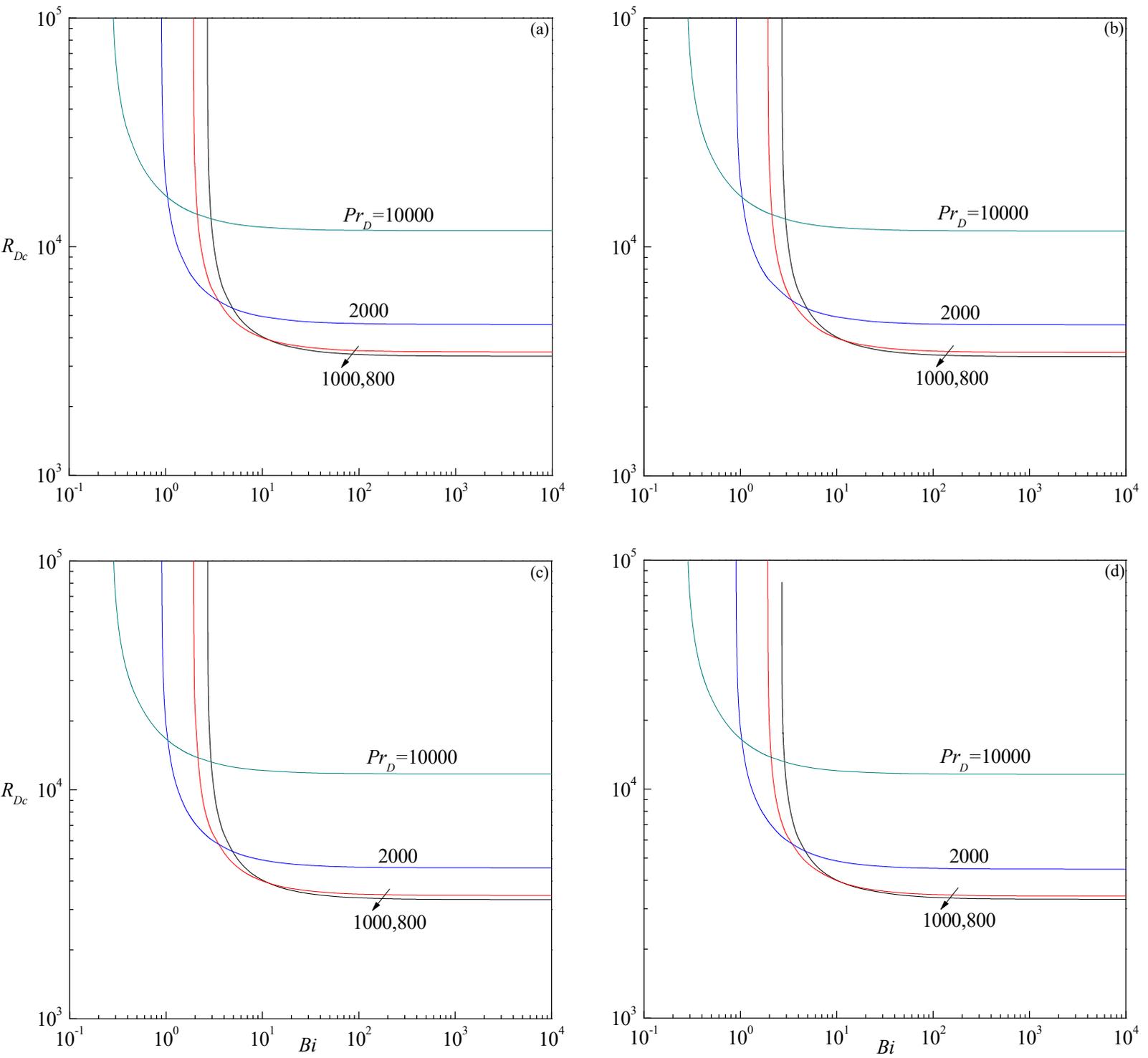

**Fig. 6.** Plots of $R_{Dc}$ versus $Bi$ for different values of $Pr_D$ and (a) $Pe=0$, (b) $Pe=1$, (c) $Pe=10$ and (d) $Pe=100$.

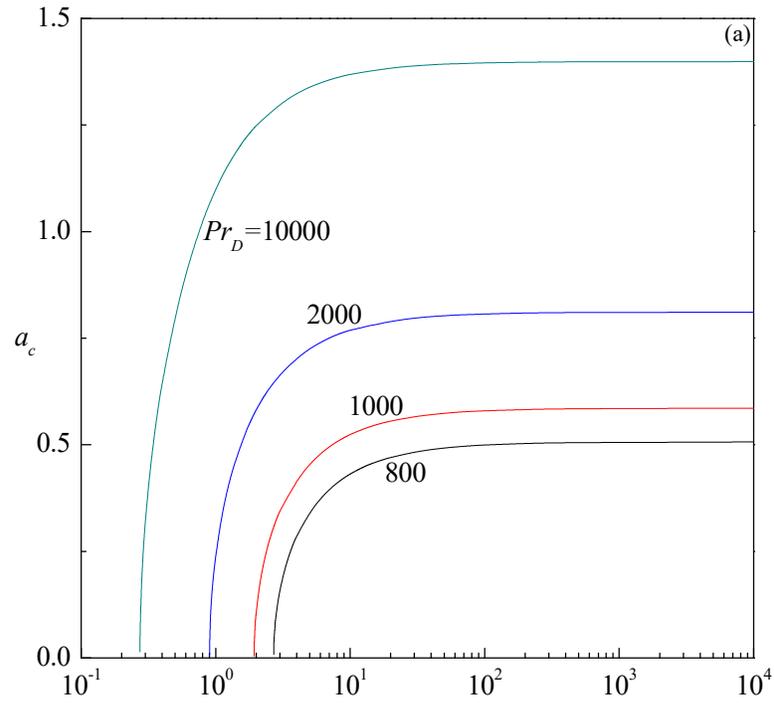
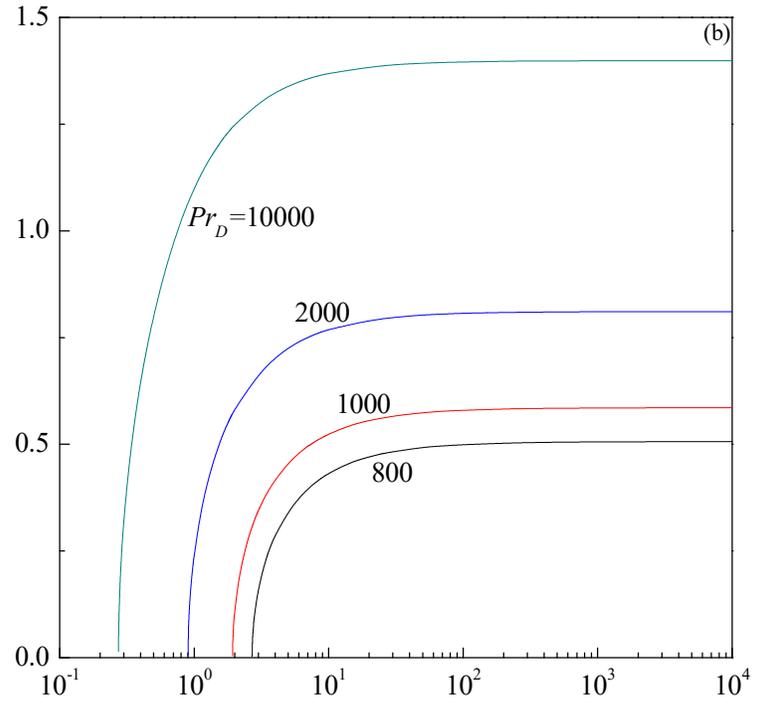
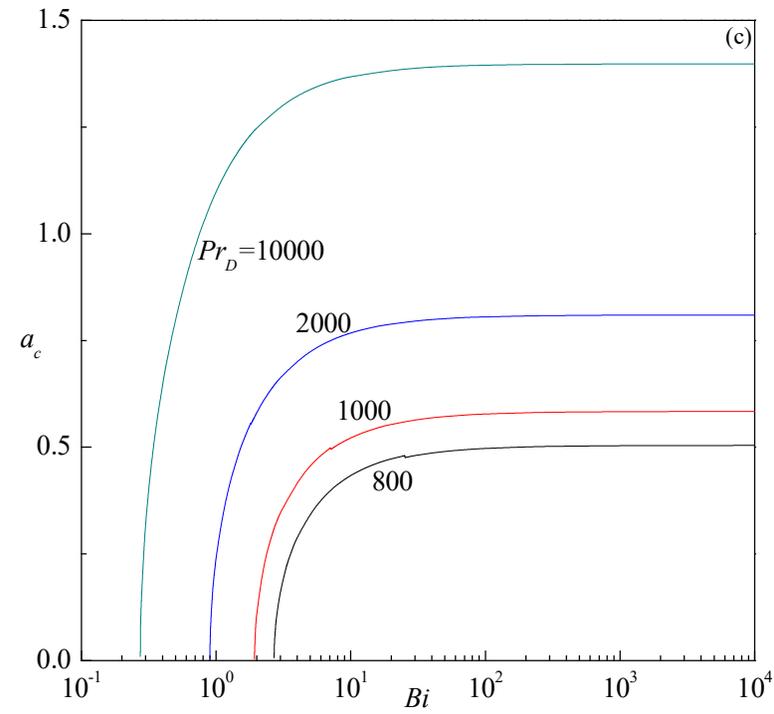
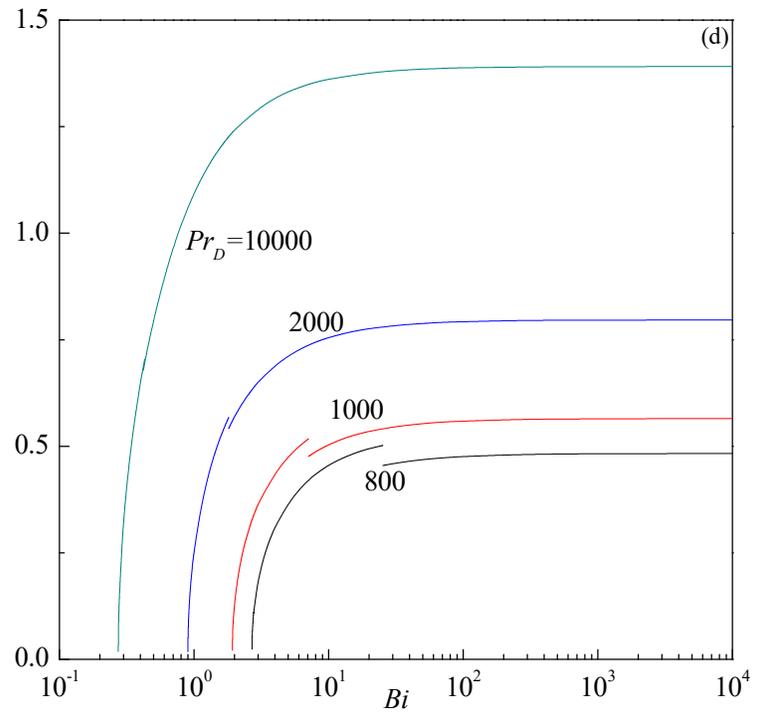

**Fig. 7.** Plots of $a_c$ versus $Bi$ for different values of $Pr_D$ and (a) $Pe=0$, (b) $Pe=1$, (c) $Pe=10$ and (d) $Pe=100$.

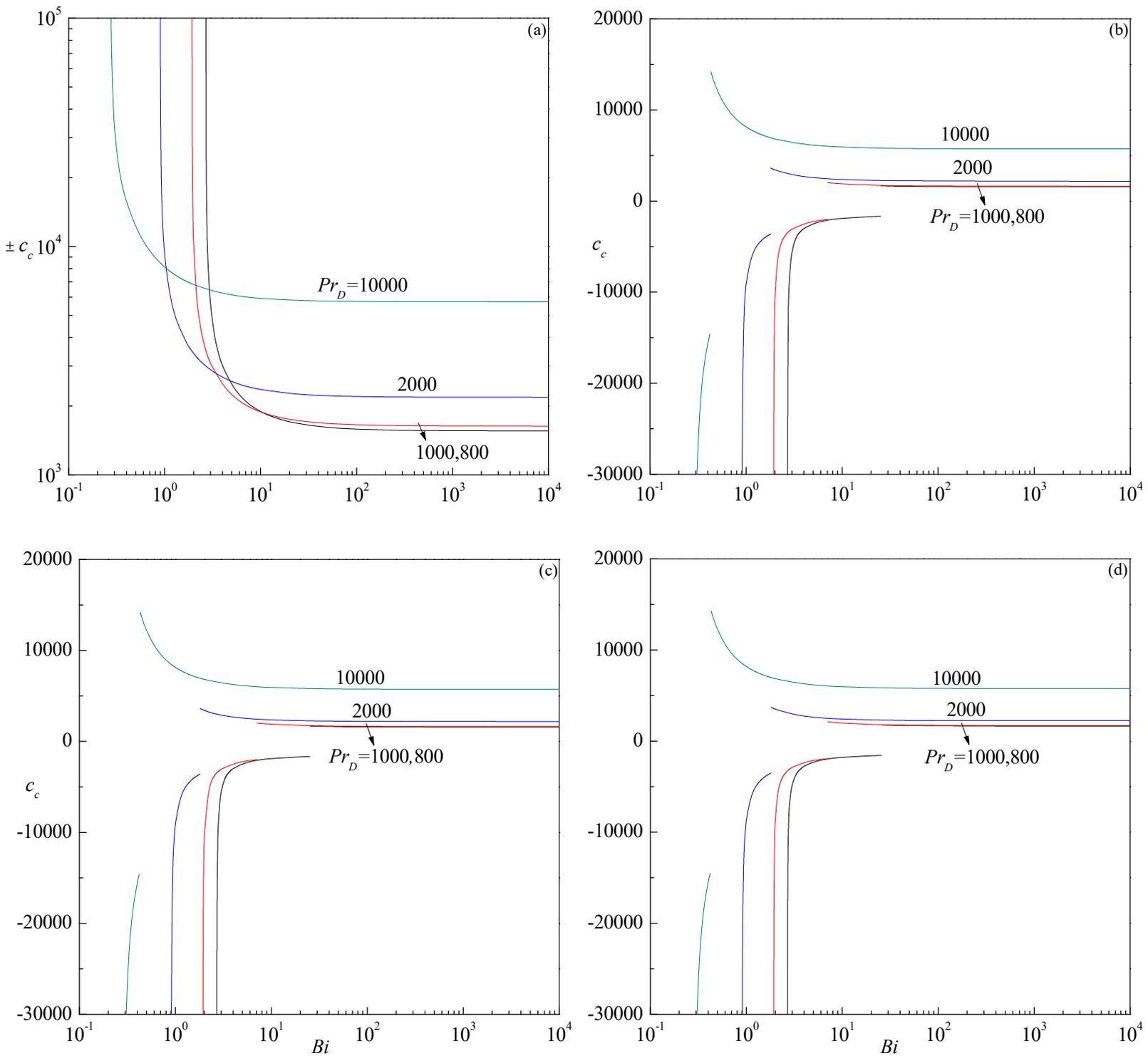

**Fig. 8.** Plots of $c_c$ versus $Bi$ for different values of $Pr_D$ and (a) $Pe=0$, (b) $Pe=1$, (c) $Pe=10$ and (d) $Pe=100$.

The corresponding critical wave number increases speedily at the initial values of $Bi$ and is poorly influenced at its higher values for all values of $Pe$ (Figs. 7a-d). The size of convection cells goes on diminishing and becomes the smallest as the Robin temperature conditions tend toward the Neumann boundary conditions that correspond to adiabatic boundaries. The effect of increasing $Pr_D$ is to reduce the size of the convection cells for all values of $Pe$. In fact, there occurs a sudden change in $a_c$ at some value of $Bi$ for each value of $Pr_D$ and for non-zero values of $Pe$ due to a shift in the mode of instability. This tendency is very insignificant at lower values of $Pe$ but visible for $Pe = 100$ (Fig. 7d). The value of $Bi$, at which the change in the mode of instability occurs, is found to decrease with increasing $Pr_D$. The trend of $\pm c_c$ closely resembles that of $R_{Dc}$ curves for $Pe = 0$ (Fig. 8a), while for other values of $Pe$ a different behavior is observed (Figs. 8b-d). For $Pe = 0$, the unstable modes travel both in the positive $z$-direction ($c_c > 0$) and negative $z$-direction ($c_c < 0$) with the same speed. For $Pe \neq 0$, $c_c$ starts with the negative sign and later it turns out to be positive with increasing $Bi$. The change of sign in $c_c$ is associated with a sudden change in $a_c$.

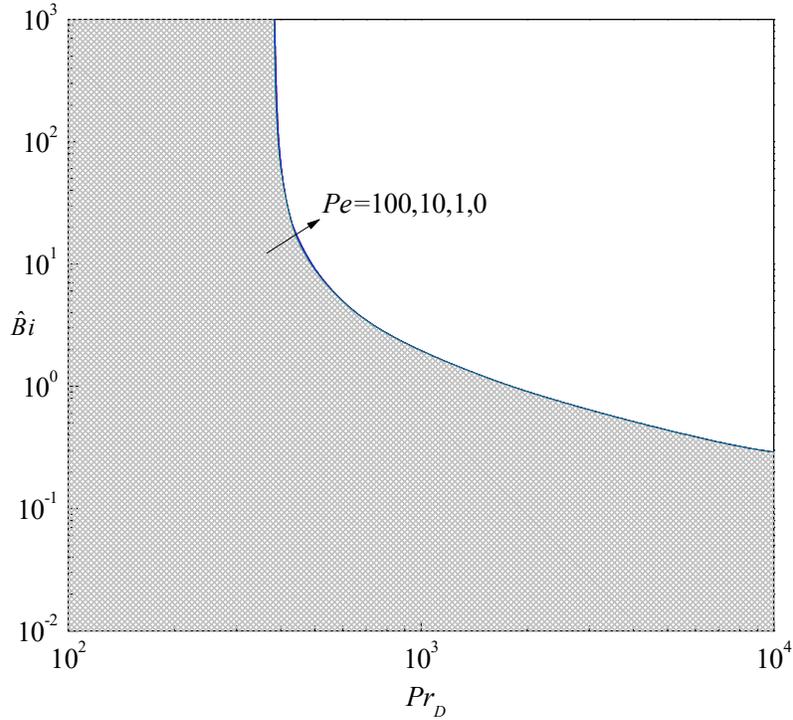

**Fig. 9.** Stability (grey) and instability (white) regions in the $(Pr_D, \hat{Bi})$-plane for different values of $Pe$.

A closer inspection of Figs. 6 and 7 discloses that $R_{Dc} \to \infty$ and $a_c \to 0$ for each considered value of $Pr_D$ when $Bi \to \hat{Bi} > 0$, where ^ stands for the transition from stability to instability. To visualize this phenomenon, the transition curves that delineate the shift from stability to instability in the $(Pr_D, \hat{Bi})$-plane are illustrated in Fig. 9 for different values of $Pe$. This figure provides a comprehensive overview of the overall stability characteristics of the base flow, offering a concise summary of its key properties. The evaluation of $\hat{Bi}$ is achieved by the solution of the stability eigenvalue problem in a large-$R_D$ range where the accuracy of the numerical solver decreases. An important result is that no instability is found when $Bi \leq \hat{Bi}$, thus a threshold to instability is found only for values of $Bi$ larger than $\hat{Bi}$. It is witnessed that $\hat{Bi}$ decreases with increasing $Pr_D$ and it is almost insignificant with increasing $Pe$.

### 6.3 Boundaries are imperfectly impermeable and isothermal ($N_0 \neq 0$ and $N_0 \to \infty$, $Bi = 0$)

Here, the impermeability boundary conditions are relaxed and a partial permeability is modelled through Robin boundary conditions for the perturbed velocity as it is more realistic when modeling a real insulation slab. Such boundaries would display an intermediate characteristic, positioning themselves between impermeable and fully permeable boundaries, striking a reasonable balance between the two extremes.

#### 6.3.1 Growth rate

| N | $Pr_D=1, Pe=1, R_D=10^3, a=1, N_0=5$ | | $Pr_D=1, Pe=100, R_D=10^3, a=1, N_0=5$ | | $Pr_D=1, Pe=1, R_D=10^3, a=1, N_0=20$ | | $Pr_D=1, Pe=1, R_D=10^4, a=1, N_0=5$ | | $Pr_D=10^3, Pe=1, R_D=10^3, a=1, N_0=5$ | |
|---|---|---|---|---|---|---|---|---|---|---|
| | $c_i$ | $c_r$ | $c_i$ | $c_r$ | $c_i$ | $c_r$ | $c_i$ | $c_r$ | $c_i$ | $c_r$ |
| 5 | -0.445013 | -0.375897 | -0.441692 | -0.224015 | -0.421365 | -0.386542 | -0.445595 | -0.384970 | -1.253595 | -201.73182 |
| 10 | -0.495917 | -0.111751 | -0.391246 | 0.013592 | -0.464316 | -0.060971 | -0.301076 | 0.000024 | -36.153326 | -66.986098 |
| 15 | -0.474233 | -0.000096 | -0.439825 | 0.047145 | -0.396345 | 0.000607 | -0.238309 | 0.001409 | -70.450240 | -240.294895 |
| 20 | -0.472781 | 0.000066 | -0.443794 | 0.045053 | -0.397212 | 0.000644 | -0.416187 | 0.000069 | -68.047615 | -236.987810 |
| 25 | -0.472763 | 0.000066 | -0.443788 | 0.045034 | -0.397176 | 0.000644 | -0.457893 | 0.000103 | -67.969650 | -236.974623 |
| 30 | -0.472764 | 0.000066 | -0.443788 | 0.045033 | -0.397177 | 0.000644 | -0.476652 | 0.000024 | -67.968675 | -236.975881 |
| 35 | -0.472764 | 0.000066 | | | -0.397177 | 0.000644 | -0.477853 | 0.000017 | -67.968674 | -236.975907 |
| 40 | | | | | -0.397177 | 0.000644 | -0.477772 | 0.000016 | -67.968674 | -236.975908 |
| 45 | | | | | | 0.000644 | -0.477780 | 0.000016 | | |
| 50 | | | | | | 0.000644 | -0.477783 | 0.000016 | | |
| 55 | | | | | | | -0.477783 | 0.000016 | | |

**Table 4.** Chebyshev approximation to the most unstable mode for different set of parameters when $Bi = 0$.

The growth rate plots $c_i$ versus $a$ are reported in Figs. 10(a-d) for different values of $Pe, R_D, N_0$ and $Pr_D$. In accordance with the recommendations provided in Table 4, the results were computed by keeping $N$ fixed at 50. The variation of $c_i$ with $a$ is somewhat contradictory in nature as $c_i$ changes its sign from negative to positive for some choices of parameters, while for higher values of $Pe$ (Fig. 10a) and for lower values of $R_D$ (Fig. 10b), $N_0$ (Fig. 10c) and $Pr_D$ (Fig. 10d) it remains always negative. Thus, the stability of the base flow is determined by the values of these parameters resulting in either a stable or unstable flow. Moreover, it is evident that the base flow is more likely to become unstable when the boundaries undergo a transition from impermeable to permeable.

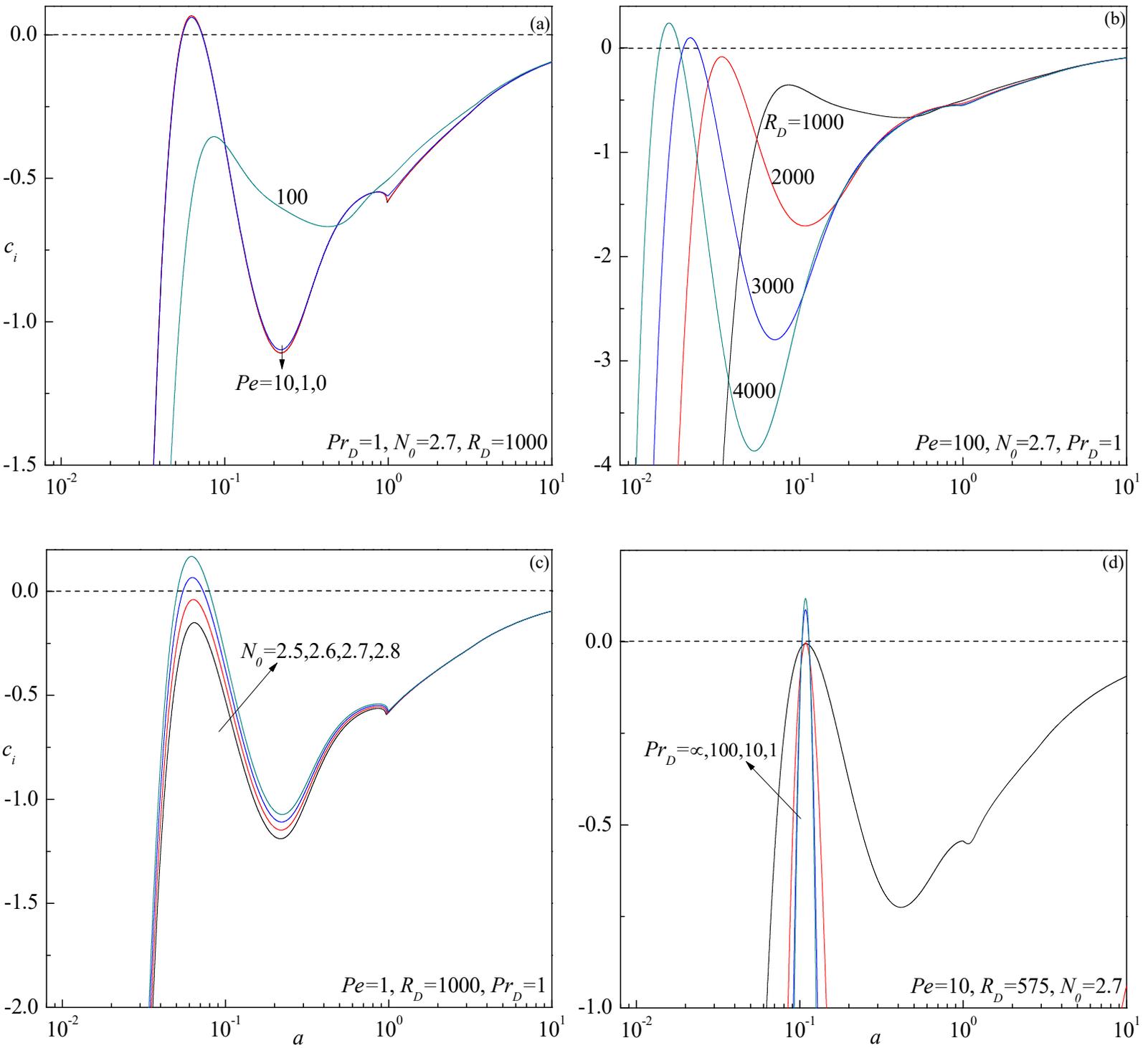

**Fig. 10.** Plots of growth rate $c_i$ versus $a$ for different values of (a) $Pe$, (b) $R_D$, (c) $N_0$ and (d) $Pr_D$.

### 6.3.2 Neutral stability curves

After conducting multiple numerical experiments to compute the critical triplets $(R_{Dc}, a_c, c_c)$ for representative values of $Pe, Pr_D$ and $N_0$, it is noticed that the maximum order of the Chebyshev polynomial may be fixed at 40 so that the computed critical stability parameters remain precise and reliable (see Table 5). The critical stability parameters $R_{Dc} = 98.540618$ and $a_c = 0.529748$ obtained under the limiting case of $Pe = 0$ (i.e., plates are stationary) and $N_0 \to \infty$ (permeable boundaries) from our numerical code are in close agreement with those of Barletta [8]. To account for the difference in the channel width, which is $2h$ in the current study instead of $h$, the obtained critical parameters were multiplied by a scaling factor of 2.

| N | $Pr_D=1, Pe=1, N_0=5$ | | | $Pr_D=1, Pe=100, N_0=5$ | | | $Pr_D=1, Pe=1, N_0=100$ | | | $Pr_D=100, Pe=1, N_0=5$ | | |
|---|---|---|---|---|---|---|---|---|---|---|---|---|
|   | $R_{Dc}$ | $a_c$ | $c_c$ | $R_{Dc}$ | $a_c$ | $c_c$ | $R_{Dc}$ | $a_c$ | $c_c$ | $R_{Dc}$ | $a_c$ | $c_c$ |
| 5  | 127.259046 | 0.398715 | 0.063428 | 414.390607 | 0.620255 | 1.130888 | 94.460620 | 0.538682 | 0.078656 | 126.891909 | 0.399356 | 0.871517 |
| 10 | 135.277551 | 0.376893 | 0.062611 | 576.619311 | 0.272764 | 1.348815 | 99.496255 | 0.518362 | 0.078471 | 134.927898 | 0.377372 | 0.869959 |
| 15 | 135.057752 | 0.377566 | 0.062626 | 574.816933 | 0.273659 | 1.348295 | 99.316414 | 0.519044 | 0.078524 | 134.708214 | 0.378045 | 0.869990 |
| 20 | 135.057735 | 0.377566 | 0.062626 | 574.804912 | 0.273676 | 1.348261 | 99.316403 | 0.519044 | 0.078524 | 134.708197 | 0.378045 | 0.869990 |
| 25 | 135.057735 | 0.377566 | 0.062626 | 574.805137 | 0.273675 | 1.348264 | 99.316403 | 0.519044 | 0.078524 | 134.708198 |   |   |
| 30 |   |   |   | 574.805137 | 0.273674 | 1.348273 |   |   | 0.078524 | 134.708198 |   |   |
| 35 |   |   |   | 574.805137 | 0.273676 | 1.348260 |   |   |   |   |   |   |
| 40 |   |   |   | 574.805137 | 0.273675 | 1.348266 |   |   |   |   |   |   |
| 45 |   |   |   | 574.805137 | 0.273675 | 1.348266 |   |   |   |   |   |   |

**Table 5.** Process of convergence of the Chebyshev collocation method when $Bi = 0$.

The neutral stability curves in the $(a, R_D)$-plane for different values of $Pe, Pr_D$ and $N_0$ are displayed in Figs. 11(a), (b) and (c), respectively. The tip of the teardrop shaped neutral stability curves in Figs. 11(a) and (c) is that there exists a maximum position where $a = a_t$ and $R_D = R_{Dt}$. Here, the subscript $t$ stands for the tip and no instability is possible for $a > a_t$. It is obvious that increasing $Pe$ (Fig. 11a) is to increase the region of stability, while an opposite trend prevails with increasing $Pr_D$ (Fig. 11b) and $N_0$ (Fig. 11c). In fact, there is no visible distinction in the neutral stability curves for $Pe = 0$ and 1, and $Pr_D$ or $N_0 = 100$ and $\infty$. From the figures, it is also evident that the instability tends to be caused by smaller wave numbers with decreasing values of $N_0$ and $Pr_D$, and increasing values of $Pe$. Moreover, the neutral stability curve for $Pe = 0$ corresponds to the stationary mode, while for all other values it corresponds to the travelling-wave mode.

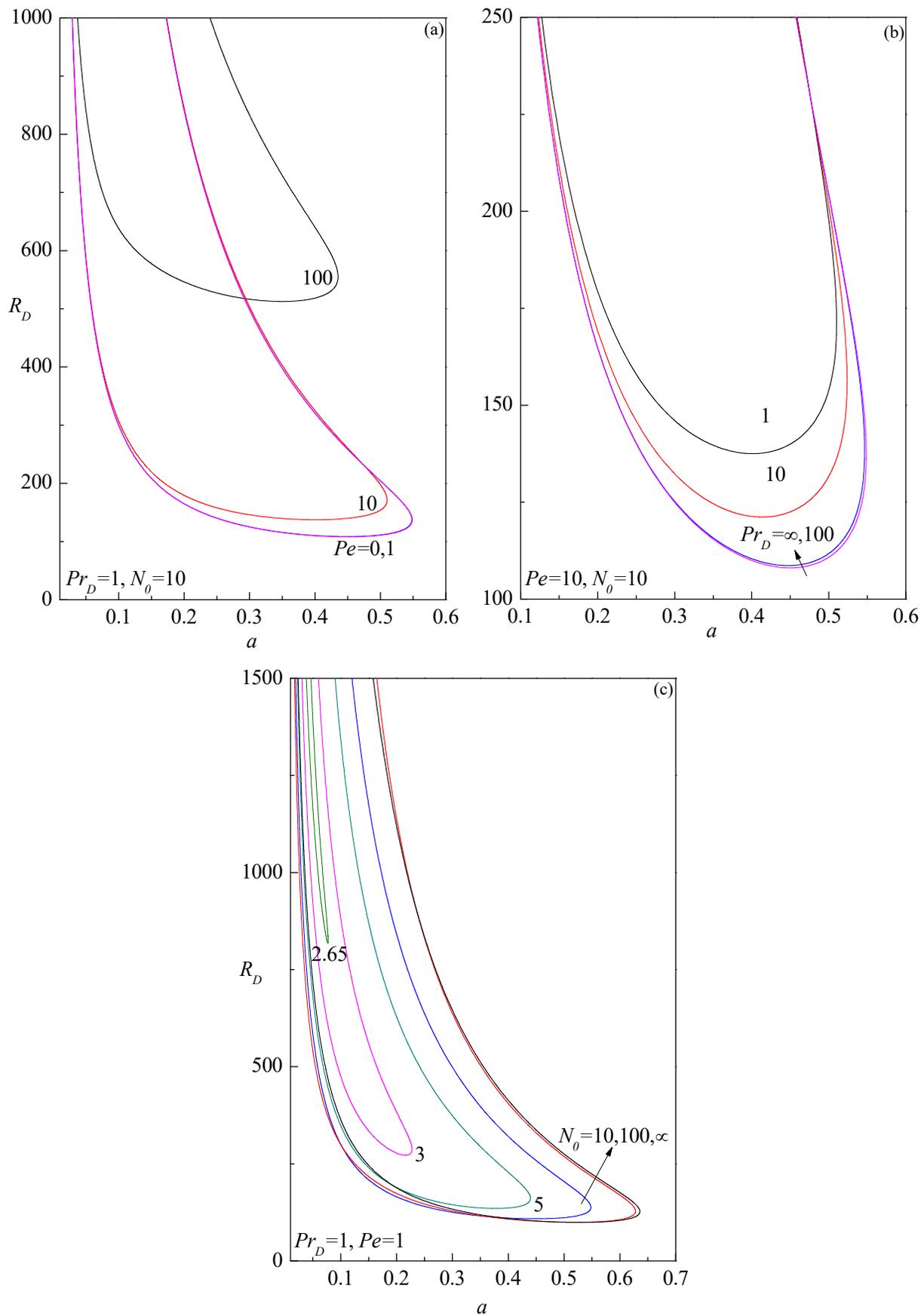

**Fig. 11.** Neutral stability curves in the $(a, R_D)$-plane for different values of (a) $Pe$, (b) $Pr_D$ and (c) $N_0$.

More interesting neutral stability curves are shown in Fig. 12(a) for different values of $N_0$ when $Pe = 100$ and $Pr_D = 10$ which are markedly different from those observed in Fig. 11. The neutral curves switchover to a bi-modal shape from a uni-modal with increasing $N_0$ and a local minimum appears in each lobe of the neutral curve for $N_0 = 200, 300$ and $400$. While the dominant mode of instability at the left-hand (right-hand) lobe of the neutral stability curve when $N_0 = 200$ (400), shifts to the right-hand (left-hand) lobe with increasing (decreasing) $N_0$. Eventually the left-hand (right-hand) lobe disappears to transform into a uni-modal shape with further increasing (decreasing) $N_0$. Though the global minimum shifts from lower to higher wave number region with increasing $N_0$, it does not affect the mode of instability as the sign of $c_c$ remains to be positive in both the minimum.

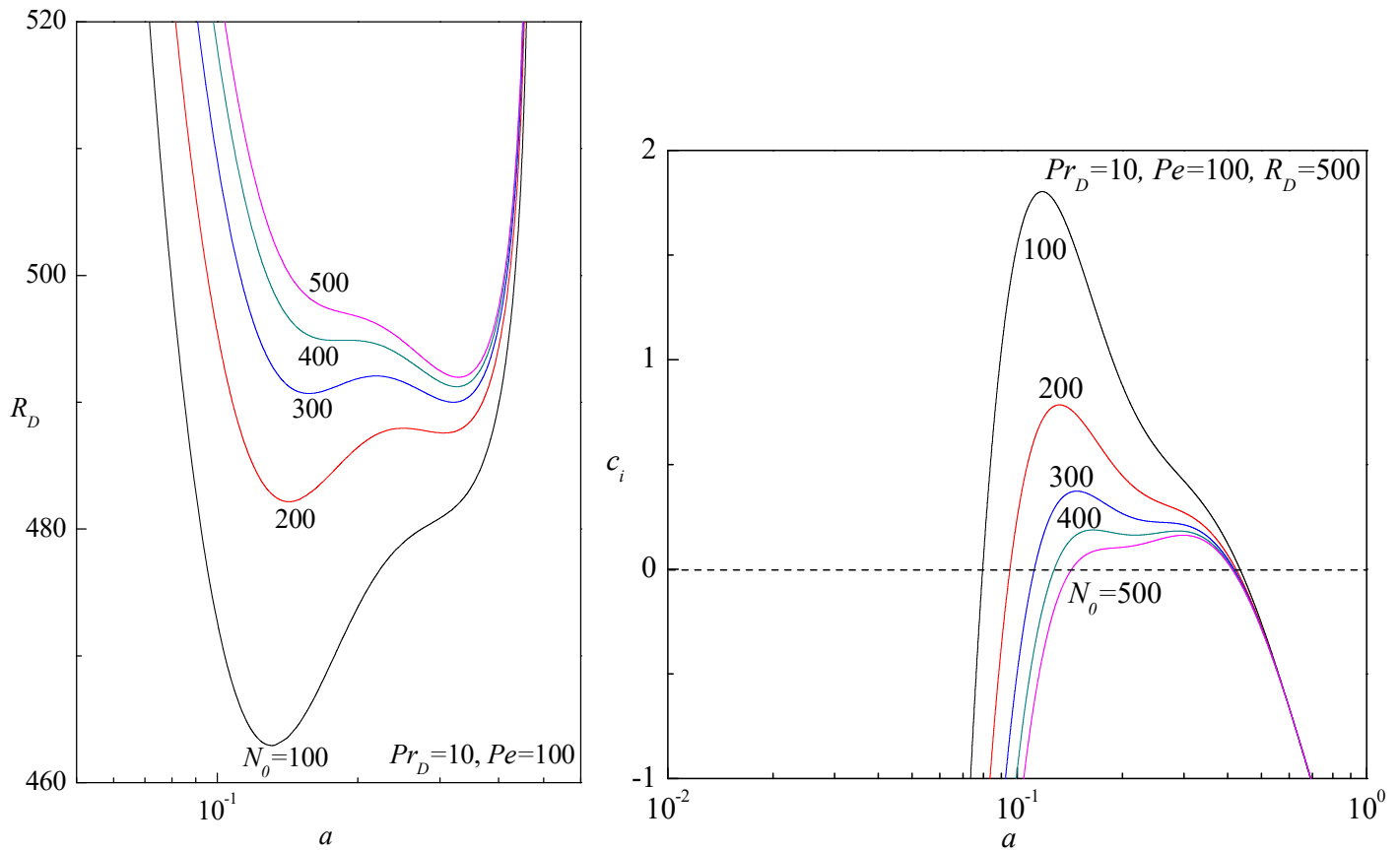

**Fig. 12.** (a) Neutral stability curves in the $(a, R_D)$-plane and (b) Plots of growth rate $c_i$ versus $a$, for different values of $N_0$.

To gain a comprehensive understanding of the neutral curves presented in Fig. 12(a), the temporal growth rate $c_i$ is computed numerically for different values of $N_0$ and illustrated in Fig. 12(b). It is noticed that double peaks appear in the temporal growth profile at higher values of $N_0$, and both peaks lie in the finite wave number regime. It is noteworthy to point out that the left peak gradually attenuates with increasing $N_0$, and ultimately vanishes but the right peak amplifies. The entire phenomena are completely consistent with the results reported in Fig. 12(a).

### 6.3.3 Condition for instability

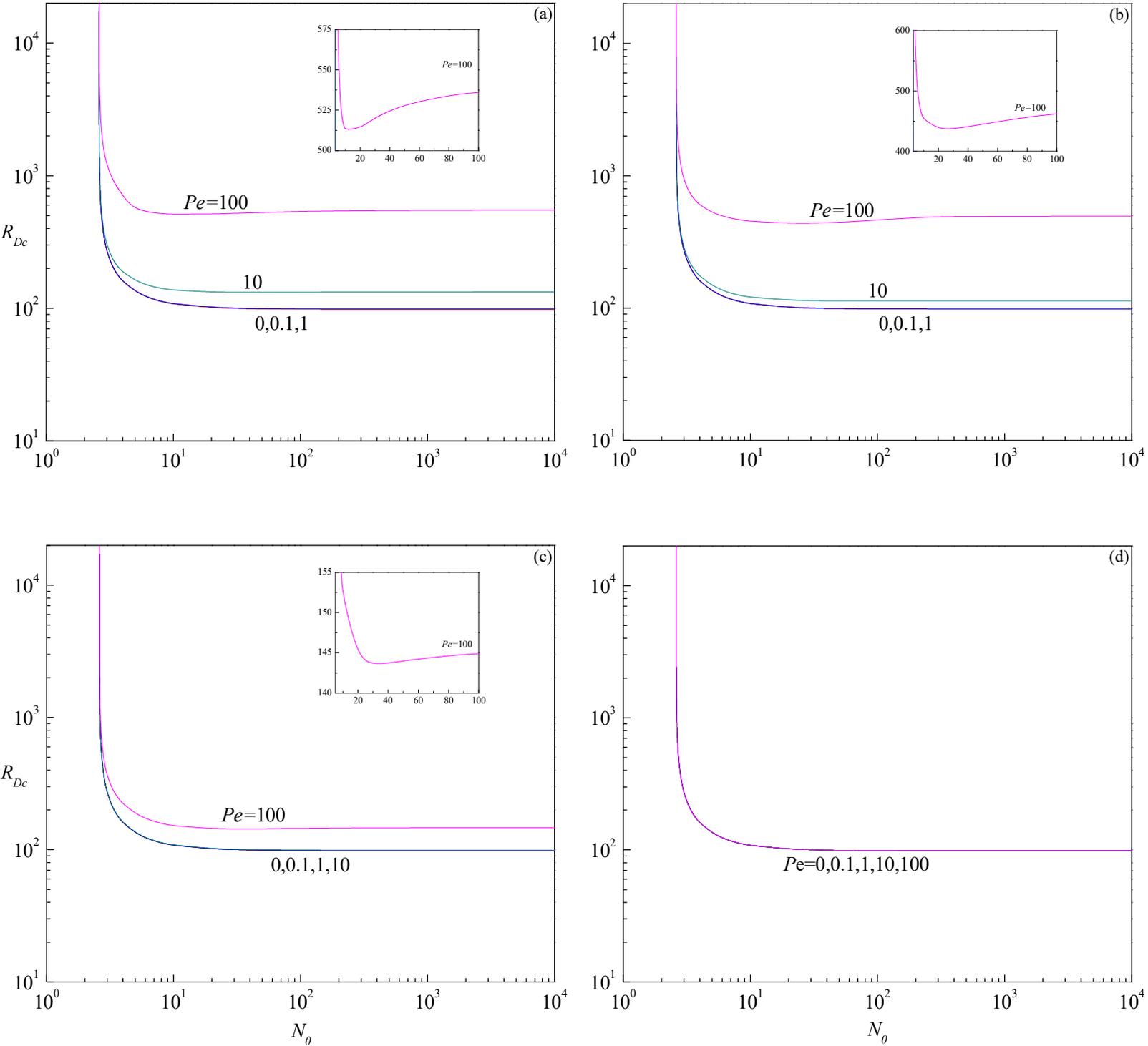

**Fig. 13.** Plots of $R_{Dc}$ versus $N_0$ for different values of $Pe$ and (a) $Pr_D=1$, (b) $Pr_D=10$, (c) $Pr_D=100$ and (d) $Pr_D\to\infty$.

The trends of critical Darcy-Rayleigh number $R_{Dc}$ versus $N_0$ are displayed for different values of $Pe$ and $Pr_D$ in Figs. 13(a-d). A departure from impermeable to permeable boundaries emerges with the change in the values of $N_0$ from $N_0 \ll 1$ to $N_0 \gg 1$. The value of $R_{Dc} \to \infty$ (i.e., the flow is stable) when $N_0 = 0 = Pe$ provides an indirect supporting argument of Rees [4] for finite values of $Pr_D$ (Figs. 13a-c) and of Gill [1] in the limit $Pr_D \to \infty$ (Fig. 13d). This tendency is found to be true for all values of $Pe$ and $Pr_D$. From the figures, it is seen that initial decline in $R_{Dc}$ followed by a sustained constancy with increasing $N_0$ for $Pe = 0, 0.1, 1$ and $10$, while for $Pe = 100$ it passes through a different minimum at $N_0 = 12.82, 23.27$ and $30.87$ for $Pr_D = 1, 10$ and $100$, respectively before attaining a constant value with further increasing $N_0$. Thus, increasing $N_0$ shows a destabilizing effect in low to moderate range of values of $Pe$ but exhibit both stabilizing and destabilizing effects on the fluid flow at higher values of $Pe$. Nonetheless, only destabilizing effect is perceived in the limit $Pr_D \to \infty$. While the curves of $R_{Dc}$ are almost overlapped when $Pr_D = 1$ for $Pe = 0, 0.1$ and $1$ (Fig. 13a), the gap between the curves reduces with increasing $Pr_D$ even for higher values of $Pe$ (Figs. 13b,c) and ultimately all the curves of different values of $Pe$ coalesce as $Pr_D \to \infty$ (Fig. 13d). Thus, the moving boundaries do not show any impact on the stability of fluid flow in the absence of time-derivative term in the momentum equation. In overall, the effect of increasing $Pe$ is to stabilize the base flow. It is also quite evident from the figures that an increase in $Pr_D$ has a destabilizing impact.

The behavior of critical wave number $a_c$ and the critical wave speed $c_c$ with $N_0$ is exhibited in Figs. 14(a-d) and Figs. 15(a-d), respectively for the corresponding parametric values chosen in Figs. 13(a-d). From Figs. 14(a-d), it is observed that the changes in the values of $a_c$ with $N_0$ is not so significant as its range of variation is very limited. In this restricted domain, $a_c$ remains unchanged at higher values of $N_0$ though it increases rapidly at its lower values. The onset of instability involves smaller and smaller wave numbers, as $a_c$ is a decreasing function of $Pe$, hence, its effect is to increase the size of convection cells, while it shows a mixed trend with increasing $Pr_D$ resulting in the both contraction and expansion of convection cells. The abrupt variation in the values of $a_c$ when $Pr_D = 10$ and $Pe = 100$ is readily noticed, and the underlying cause for this behavior is clearly confirmed by the curves displayed in Fig. 12. From the figures, it is also evident that the size of convection cells of fixed walls is smaller than the moving walls. From Figs. 15(a-d), it is seen that $c_c$ increases gradually and becomes independent of $N_0$ for all values of $Pr_D$ and $Pe$ except when $Pr_D = 1$ and $10$ with $Pe = 100$ (Figs. 15a,b). As $Pe$ increases, $c_c$ increases in a manner similar to that of basic velocity. The effect of increasing $Pr_D$ is to increase $c_c$. It is noted that the instability is through travelling-wave modes in the case of $Pe \neq 0$ and that too only positive $c_c$ exists implying that unstable modes travel only in the upward direction. For $Pe = 0$, stationary disturbances govern the onset of instability.

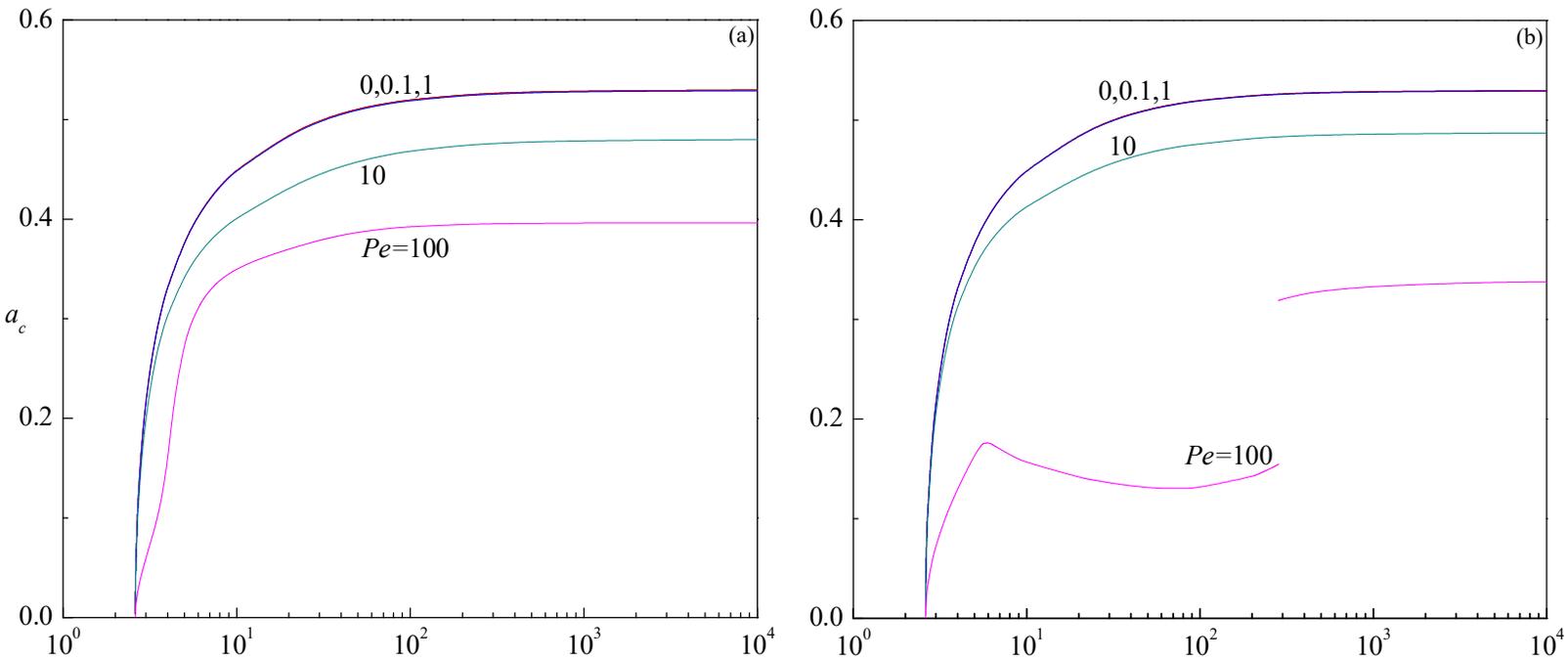

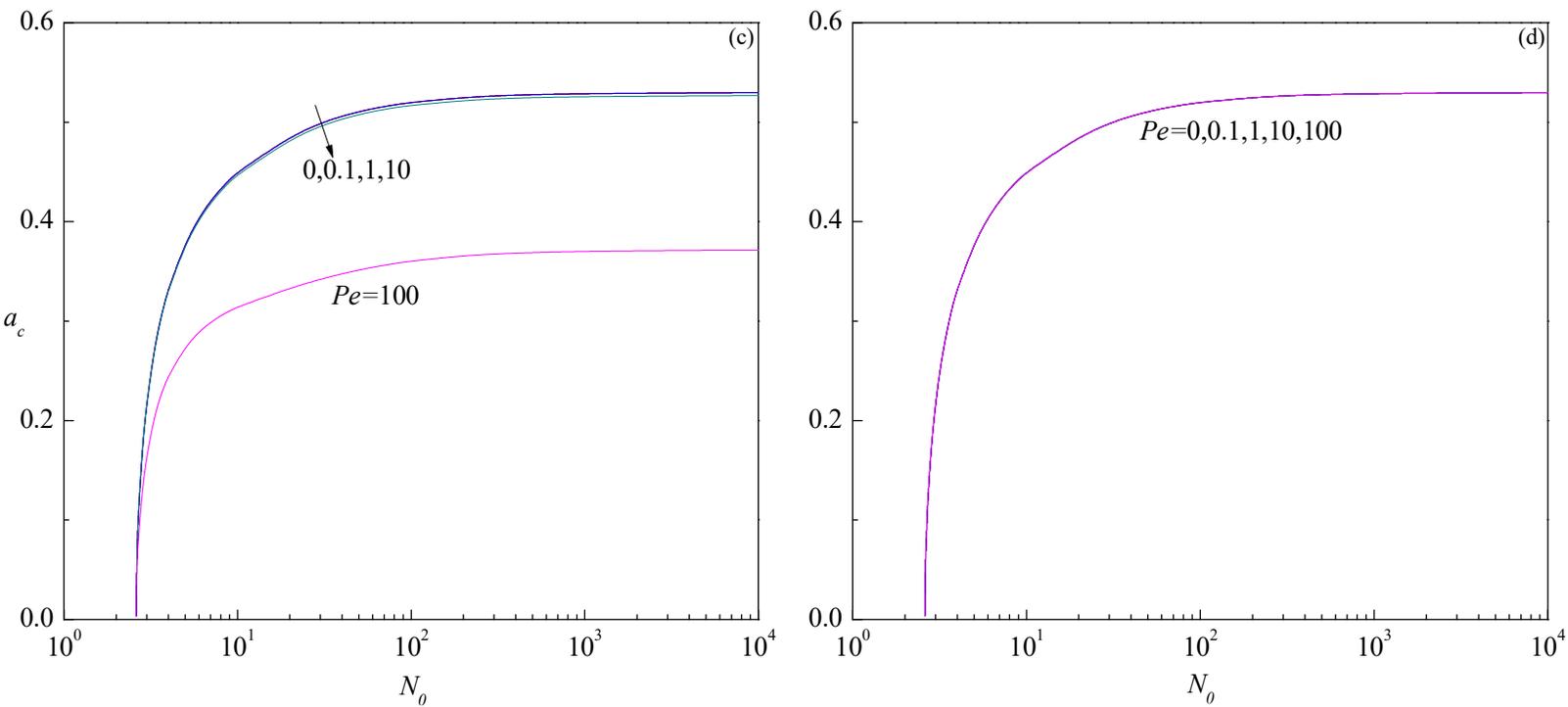

**Fig. 14.** Plots of $a_c$ versus $N_0$ for different values of $Pe$ and (a) $Pr_D = 1$, (b) $Pr_D = 10$, (c) $Pr_D = 100$ and (d) $Pr_D \to \infty$.

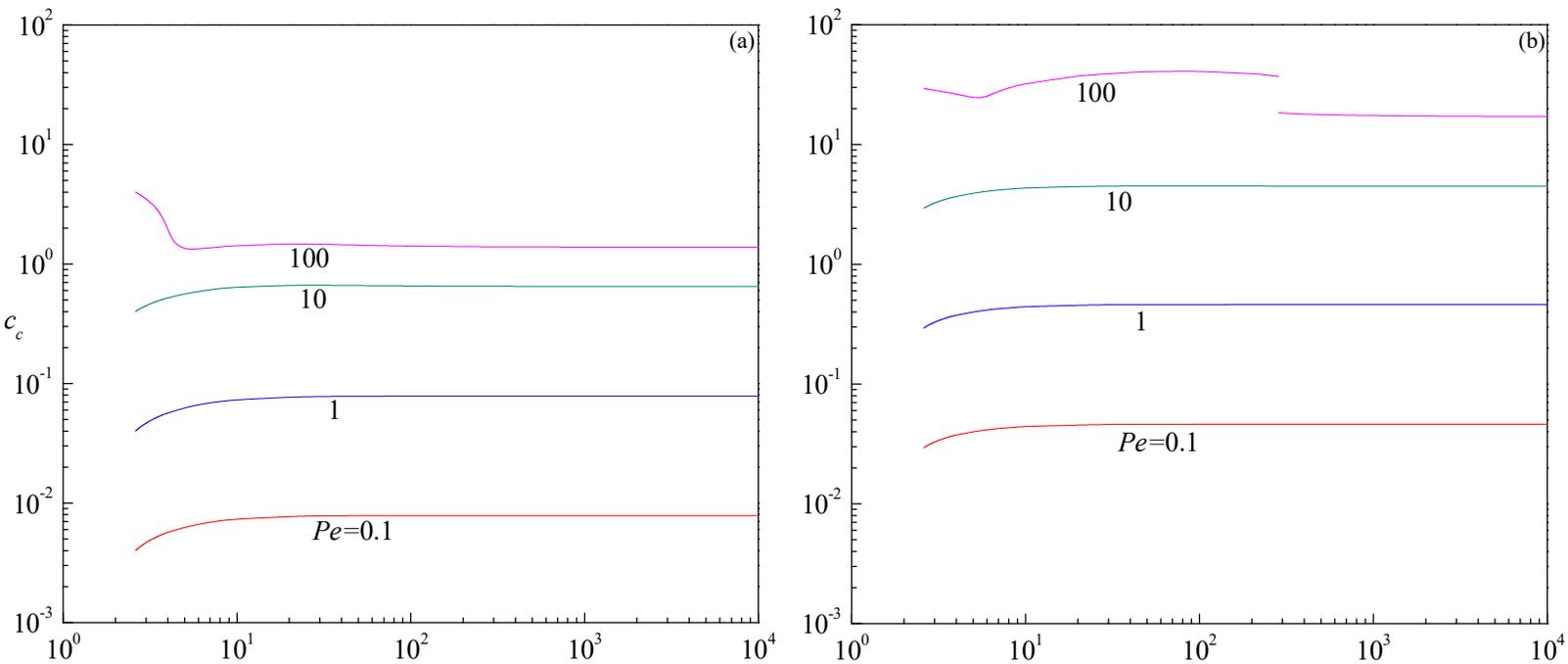

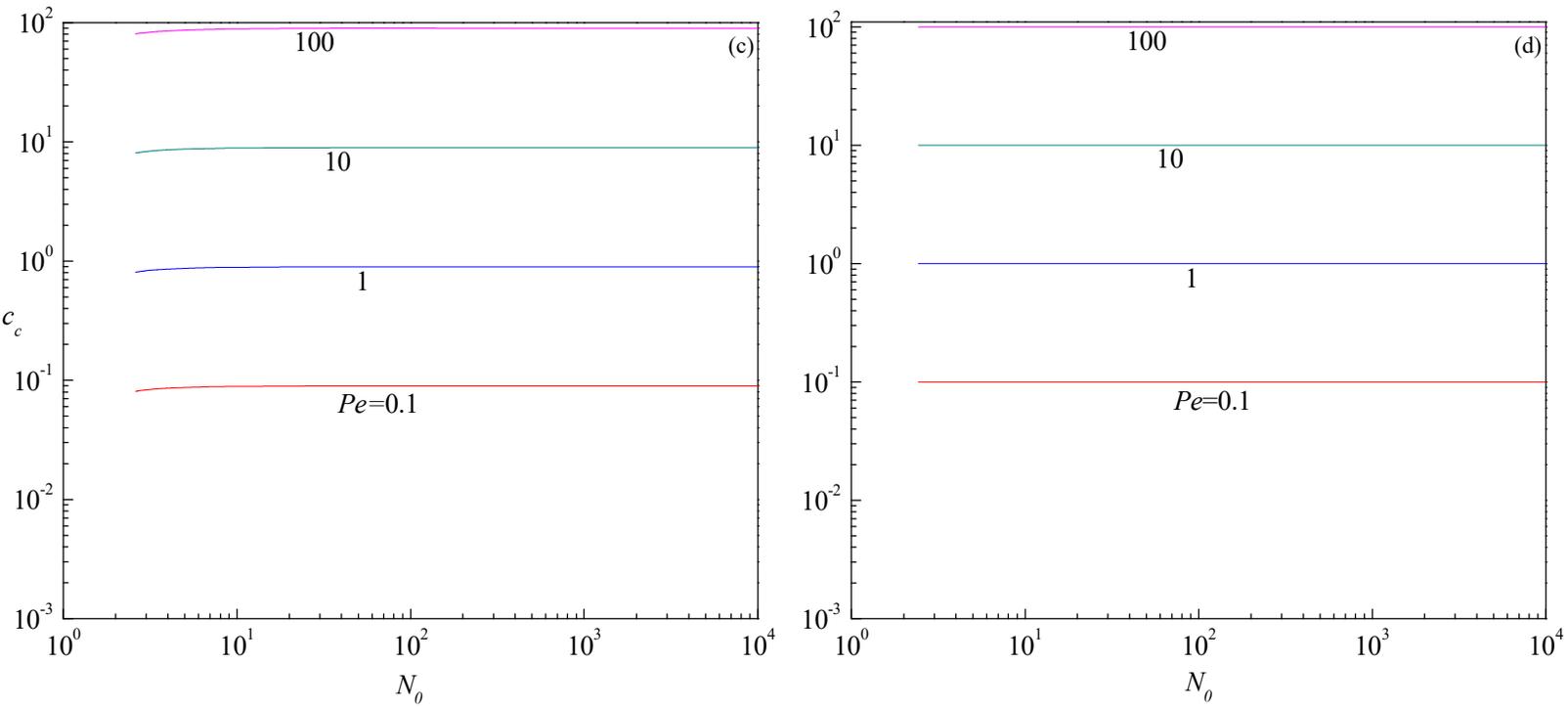

**Fig. 15.** Plots of $c_c$ versus $N_0$ for different values of $Pe$ when (a) $Pr_D=1$, (b) $Pr_D=10$, (c) $Pr_D=100$ and (d) $Pr_D \to \infty$.

The violation of impermeability condition leads to the instability of the basic parallel flow as noticed in Figs. (13-15). To comply with this, the precise values of $N_0$ are computed for different values of $Pe$ and $Pr_D$ at which $R_{Dc} \to \infty$ and also $a_c \to 0$, called the threshold value of $N_0$, say $\hat{N}_0$. It is found that the transition from stability to instability occurs when $\hat{N}_0 \geq 2.610$, independent of $Pe$ and $Pr_D$, and the same is evident from Fig. 16.

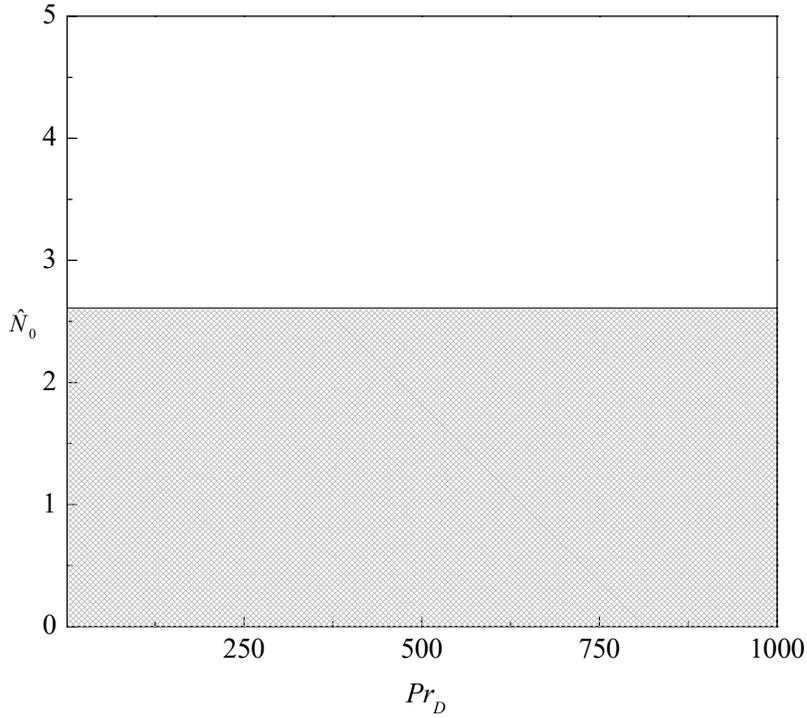

**Fig. 16.** Stability (grey) and instability (white) regions in the $(Pr_D, \hat{N}_0)$-plane for different values of $Pe$.

### 6.4 Boundaries are imperfectly impermeable and imperfectly conducting $(N_0 \neq 0, Bi \neq 0)$

It has been established in the previous sections that the base flow becomes unstable if the boundaries are imperfectly impermeable and isothermal or impermeable and imperfectly conducting. In light of these findings, the consideration of general boundary conditions on velocity and temperature perturbations seems to be pragmatic as they can reveal unforeseen and remarkable results.

### 6.4.1 Neutral stability curves

The neutral stability curves are reported in Figs. 17(a-d) for different values of $Bi, N_0, Pr_D$ and $Pe$. The effect of $Bi$ and $N_0$ is destabilizing since the curves drift downwards as these parameters increase. As $Pr_D$ increases for the considered sets of parameters, the minimum value of $R_D$ also experiences an increase. The wave speed remains to be positive along the entire curve when $Pe = 1$ (Figs. 17a, b). For $Pe = 100$, however, little tilting in the neutral curves could be seen at some values of $a$ (Figs. 17c, d). This is due to the change in the sign of $c$ from a positive to a negative value with increasing $a$, but they do not impact the onset of instability.

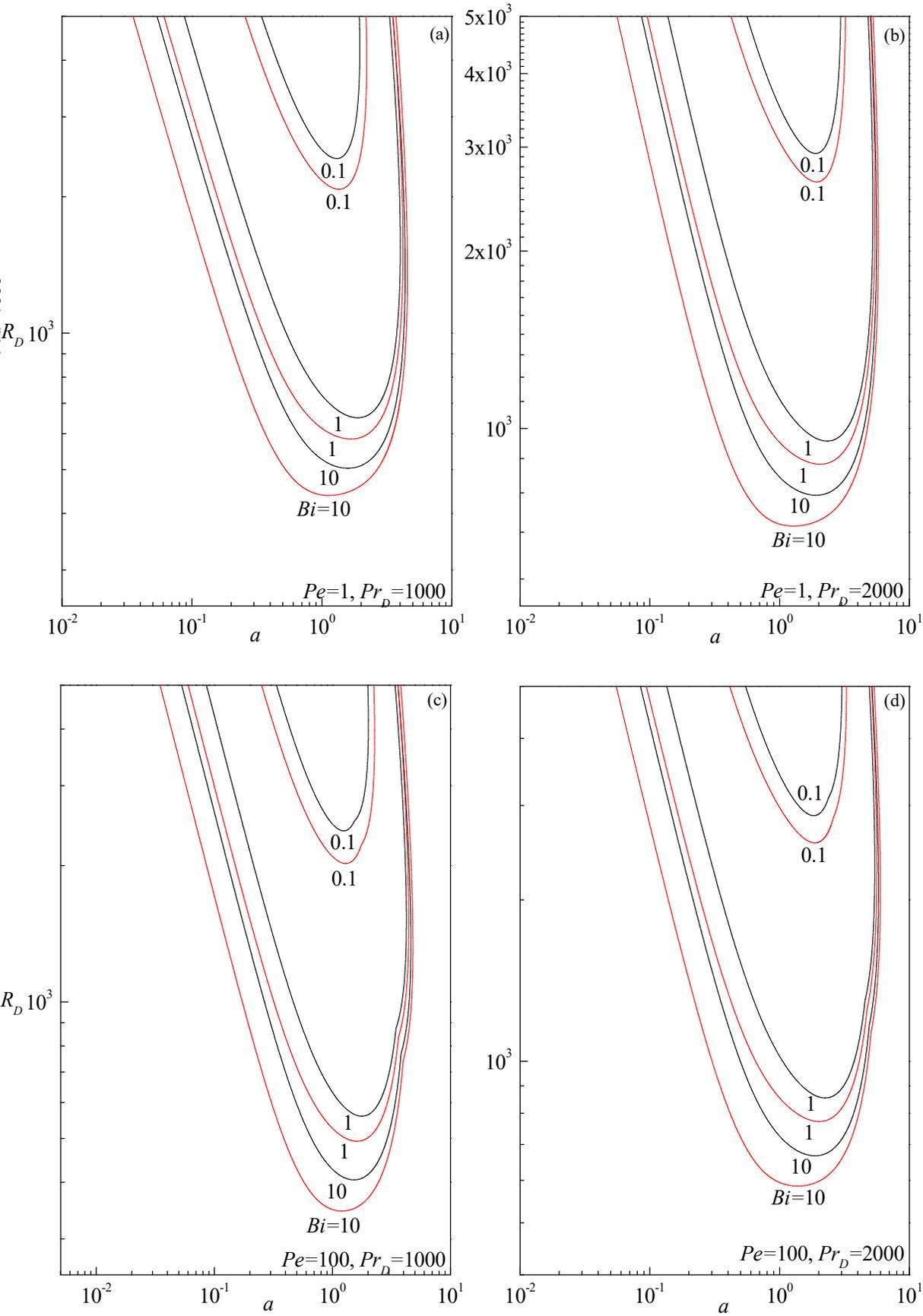

**Fig. 17.** Neutral stability curves in the $(a, R_D)$-plane for different values of $Bi, N_0, Pe$ and $Pr_D$. The black and red color lines respectively represent $N_0 = 1$ and 2.

Figures 18(a-f) exemplify the evolution of neutral stability curves as $Bi$ increases when $N_0 = 3$, $Pr_D = 1000$ and $Pe = 1$. Some novel outcomes that were not observed in the previous two cases are found in a certain range of values of $Bi$. Figure 18(a) shows the result for $Bi=0.001$ and note that only one branch of the neutral curve turns up, representing the travelling-wave, and displaying a parabolic shape. When $Bi = 0.02$, another neutral curve exists in the subsequent wave number regime, featuring a higher minimum that corresponds to the same mode (Fig. 18b). As the value of $Bi$ continues to increase, the neutral curve possessing lower (higher) minimum gets diminished (enlarged) towards the lower (higher) wave number side. Concurrently, the curve corresponding to the lower wave number region shifts the global minimum of $R_D$ both downwards and upwards as presented in Figs. 18(c-e). Ultimately, this neutral curve disappears leaving behind only the single neutral curve positioned at the higher wave number side (Fig. 18f).

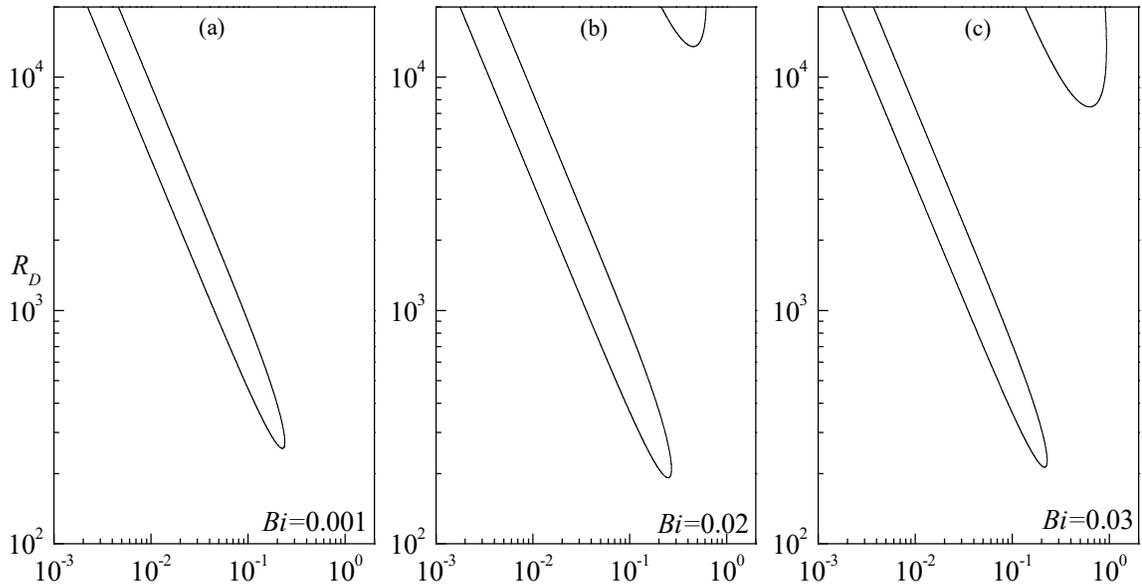

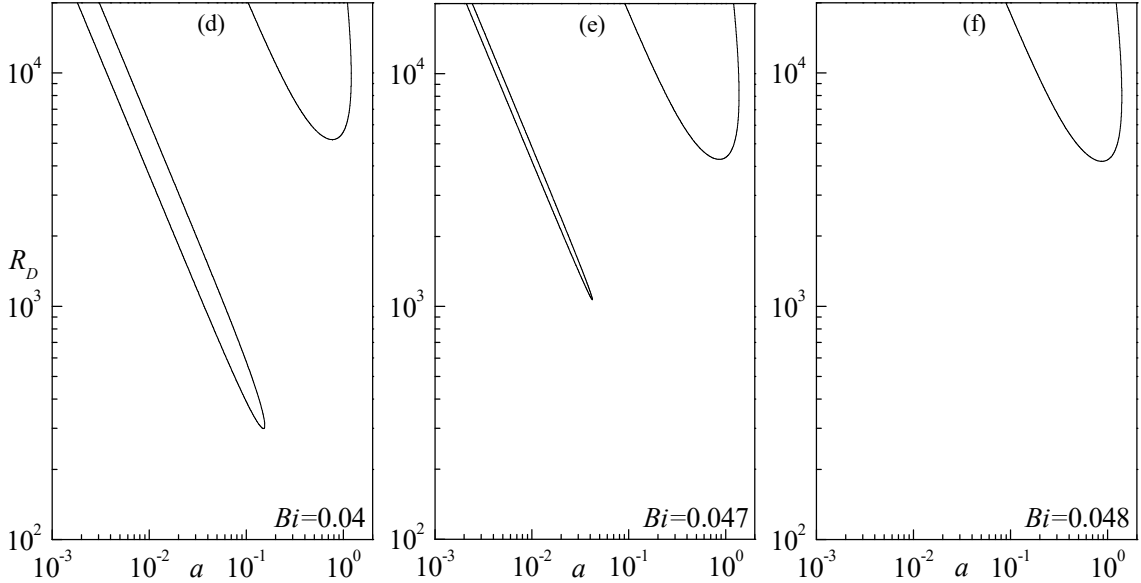

**Fig. 18.** Evolution of neutral stability curves for different values of $Bi$ when $N_0 = 3, Pe = 1$ and $Pr_D = 1000$.

### 6.4.2 Condition for instability

Figures 19(a) and (b) depict the variation of $R_{Dc}$ with respect to $Bi$ for two different values of $Pe$ specifically $Pe = 1$ and 100, wherein the results are presented for different values of $N_0$ and $Pr_D$. Notably, the findings significantly diverge from those of prior cases due to the emergence of two different branches of neutral stability curves depending upon the specific combinations of $Bi$ and $N_0$. For the case where $N_0 = 1$, instability is observed to manifest only beyond a certain threshold value of $Bi$, similar to observations in preceding cases. In this scenario, as $Bi$ increases, $R_{Dc}$ exhibits a continuous decline. Conversely, for the cases where $N_0$ takes values of 3, 5, and 10, instability is found to be feasible across the entire range of $Bi$. However, in these instances, $R_{Dc}$ showcases two distinct regimes distinguished by the values of $Bi$, arising from the presence of dual branches of neutral stability curves, which are disconnected and having a different minimum (as exemplified in Figure 18). A notable transition in the minimum value of $R_D$ from a lower wave number region (initial regime) to a higher one (latter regime) takes place at specific $Bi$ values: 0.0476, 0.0956, and 0.1369 for $N_0 = 3$, 5 and 10, respectively. These transitions are insensitive to the values of $Pr_D$ and $Pe$. In the pre-transition phase (initial regime), $R_{Dc}$ exhibits independence from the parameter $Pr_D$, yet it assumes a minimum value at $Bi$ =0.02, 0.03, and 0.05 for $N_0 = 3$, 5 and 10, respectively, indicating a conflicting behavior. Upon surpassing these critical $Bi$ values (entering the latter regime), $R_{Dc}$ showcases a decreasing trend with increasing $Bi$ for both the values of $Pr_D$. Although the impact of increasing $Pr_D$ introduces conflicting trends at lower $Bi$ values (beginning of latter regime), it ultimately stabilizes the flow as $Bi$ increases further. Upon closer examination of the figures, it becomes evident that the presence of $Pe$ expedites the onset of instability. Consequently, the manipulation of boundary characteristics offers a viable avenue for the control of fluid flow instability.

The corresponding values of $a_c$ and $c_c$ are presented in Figs. 20 (a,b) and 21(a,b), respectively. In the initial regime of $Bi$, $a_c$ displays a slight increase but swiftly gives way to a pronounced decrease, ultimately converging toward zero as $Bi$ increases, particularly evident for $N_0 = 3$, 5 and 10. Upon transitioning into the later $Bi$ range, $a_c$ experiences a sudden ascent, leading to a significant alteration in the value of $R_{Dc}$. Within this regime, $a_c$ demonstrates a characteristic pattern where it attains a maximum before stabilizing at elevated $Bi$ values, a trend consistent across all considered $N_0$ values. Consequently, the dimension of the cellular structure exhibits variations, either expanding or contracting, dependent upon the interplay of $N_0$, $Bi$ and $Pe$, with a general tendency to diminish as $Pr_D$ increases. The behavior of the critical wave speed $c_c$, presents a paradoxical nature as $Bi$ increases, showcasing diverse propensities for distinct combinations of $N_0$ and $Pr_D$ when $Pe = 1$ and 100. Additionally, a transition in the sign of $c_c$ is evident, progressing from negative to positive as $Bi$ increases for cases where $N_0 = 1$. This shift signifies a transformation in cell's motion, shifting from a downward trajectory to an upward one. Conversely, for $N_0 = 3$, 5 and 10, $c_c$ consistently maintains a positive value across the entire $Bi$ spectrum. Furthermore, the critical wave speed experiences augmentation with rising values of $Pe$ and $Pr_D$.

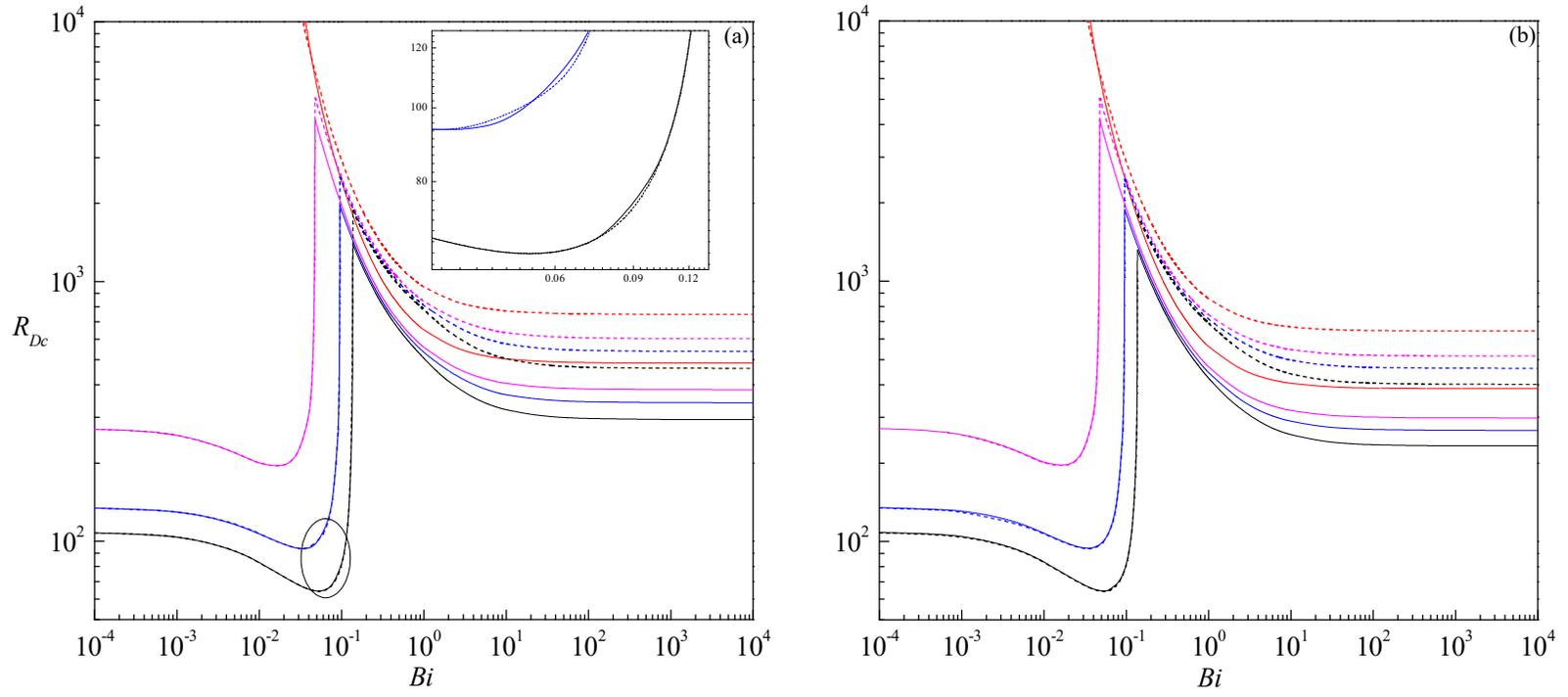

**Fig. 19.** Plots of $R_{Dc}$ versus $Bi$ for different values of (a) $Pe=1$ and (b) $Pe=100$. The solid and dashed lines respectively represent $Pr_D = 1000$ and $2000$. The red, magenta, blue and back, color lines respectively represent $N_0 =1, 3, 5$ and $10$. This convention also applies to Figs. 18 and 19.

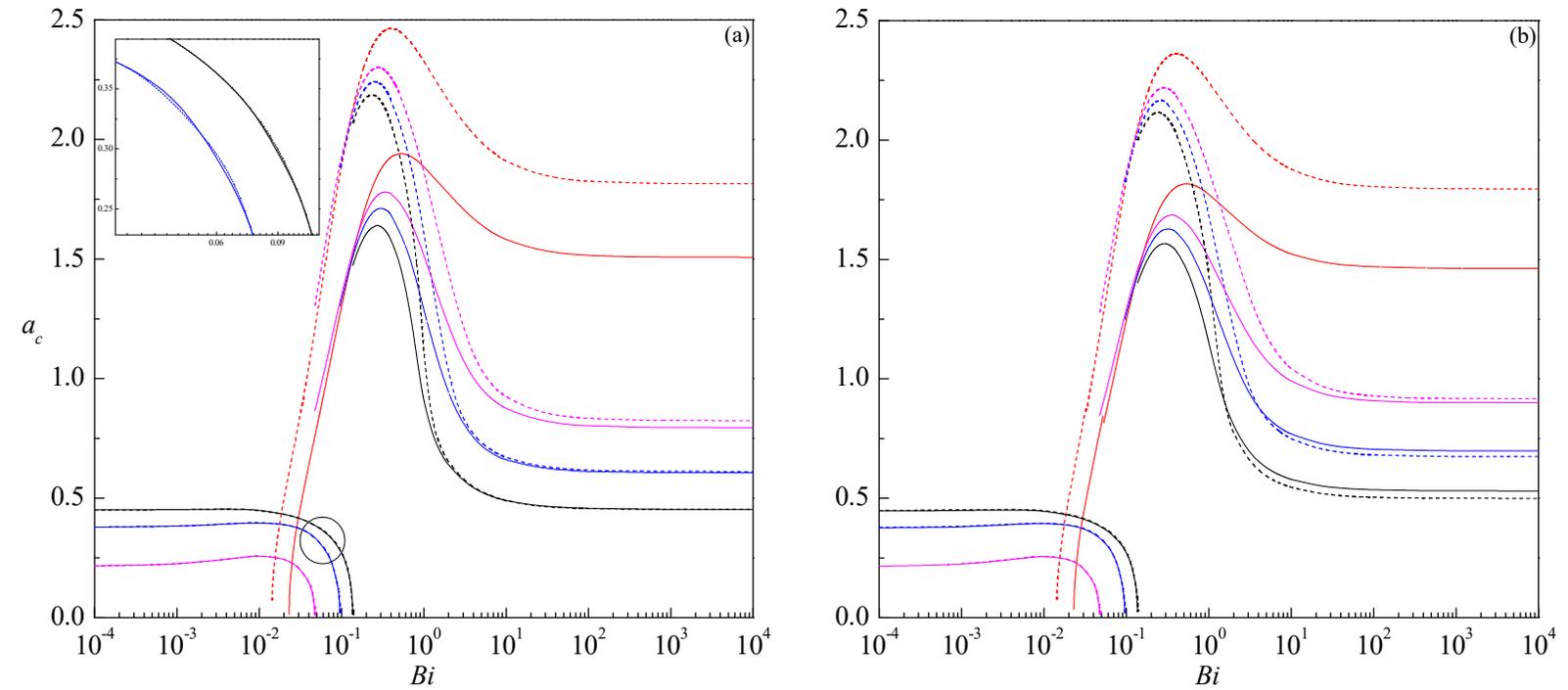

**Fig. 20.** Plots of $a_c$ versus $Bi$ for different values of (a) $Pe=1$ and (b) $Pe=100$.

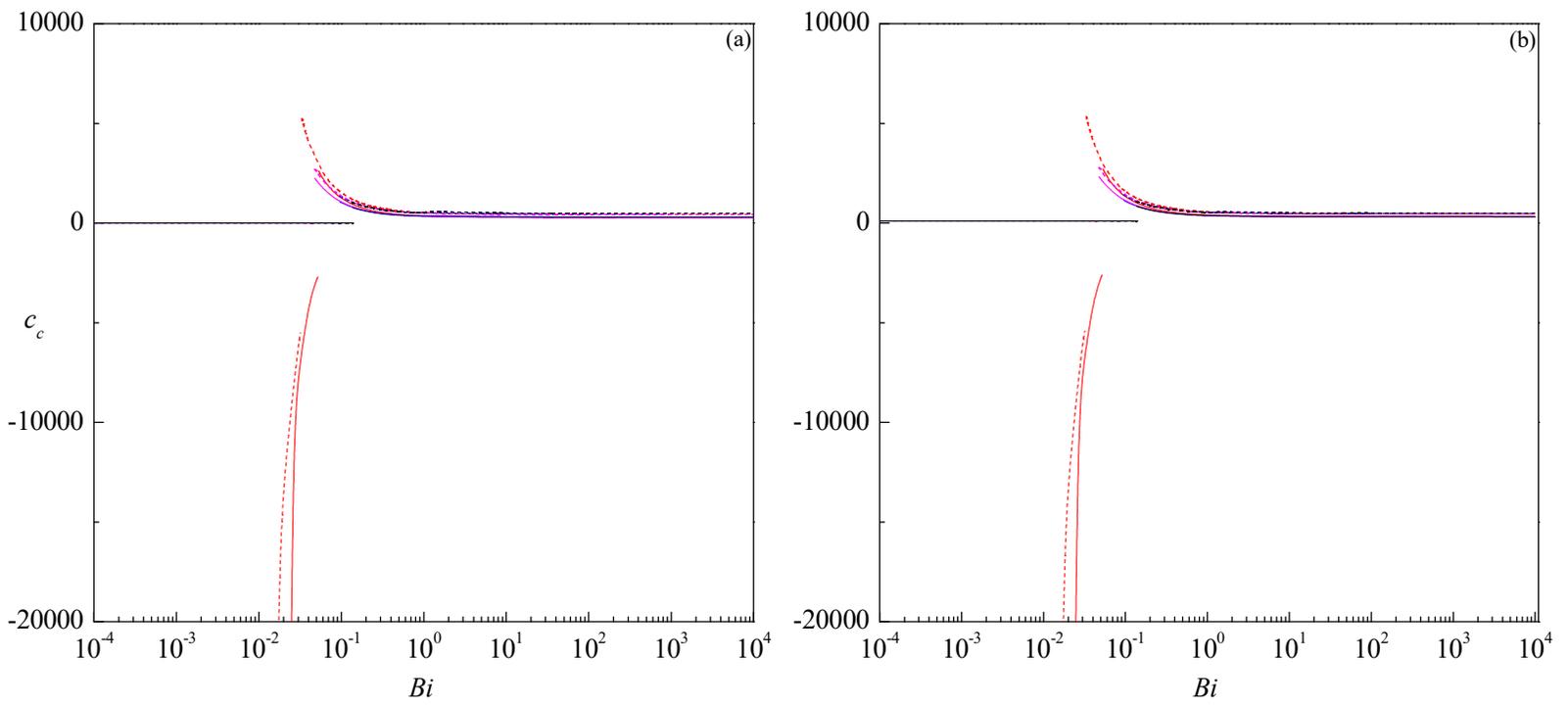

**Fig. 21.** Plots of $c_c$ versus $Bi$ for different values of (a) $Pe=1$ and (b) $Pe=100$.

**6.5 Boundaries are permeable and adiabatic** $(N_0 \to \infty \leftarrow Bi)$

For the boundaries considered, instability emerges without any restriction on the physical parameters unlike in the previous cases. Figure 22(a) shows the variation of $R_{Dc}$ as a function of $Pe$ for different values of $Pr_D$. It is evident that for lower values of $Pr_D$, $R_{Dc}$ is a decreasing function of $Pe$. However, as $Pr_D$ increases, $R_{Dc}$ remains constant till a specific $Pe$ value, denoted as $\tilde{Pe}$, and subsequently decreases monotonically as $Pe$ continues to rise. Moreover, $\tilde{Pe}$ increases with higher $Pr_D$ values. Furthermore, one may note that the dependence of $Pe$ becomes weaker and weaker as $Pr_D$ increases since $R_{Dc}$ becomes independent of $Pe$ for $Pr_D = 10^4$. Conversely, for a fixed $Pe$ value, $R_{Dc}$ increases with increasing $Pr_D$. Thus, the departure from infinitely Prandtl-Darcy case hastens the onset of convective instability. The corresponding wave number is displayed in Fig. 22(b). It is observed that the critical wave number initially increases and then decreases as $Pe$ increases for $Pr_D=1$, while for other values of $Pr_D$, $a_c$ increases with increasing $Pe$. The unstable modes exclusively propagate in the upward direction, as they are contingent upon the existence of positive $c_c$ solely.

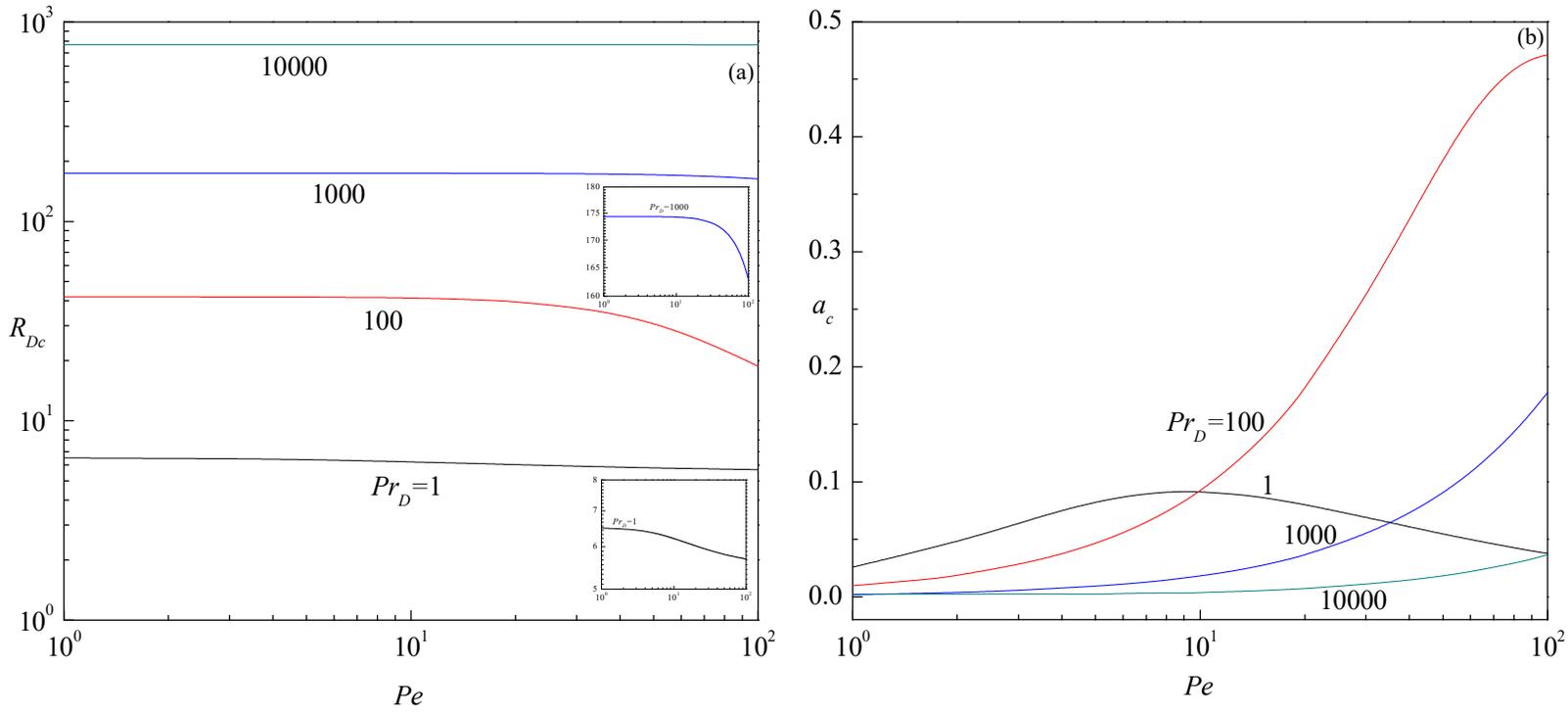

**Fig. 22.** Plots of (a) $R_{Dc}$ and (b) $a_c$ versus $Pe$ for different values of $Pr_D$ when $N_0 \to \infty \leftarrow Bi$.

## 7. Conclusions

The investigation centers on the stability analysis of natural convection in a fluid-saturated vertical porous slab. The study considers the relative motion of boundaries and incorporates Robin-type conditions on perturbation velocity and temperature fields. The underlying dynamics of convection is governed by Darcy's law encompassing a time-dependent velocity term. The thermal buoyancy effect is captured through the Oberbeck-Boussinesq approximation. The basic flow has been determined analytically, revealing that the interplay of boundary motion significantly influencing this flow. Without any loss of generality, a two-dimensional linear stability analysis has been carried out instead of a three-dimensional case to assess the critical conditions for the onset of instability. The characterization of velocity and temperature boundary conditions hinges on the parameters $N_0$ and the Biot number $Bi$, both of which fall within the range $[0, \infty]$. The boundaries are assumed to be impermeable/isothermal if $N_0 = 0/Bi = 0$, perfectly permeable/adiabatic if $N_0 \to \infty / Bi \to \infty$, while the boundaries are categorized as partially permeable/conducting for the intermediate values of $N_0 / Bi$. Crucial parameters impacting the stability of the fluid flow include the Prandtl-Darcy number $Pr_D$, the Péclet number $Pe$ and the Darcy-Rayleigh number $R_D$. To ascertain stability, a numerical solution is obtained for the stability eigenvalue problem. This is achieved through the application of the Chebyshev collocation method, which is employed for diverse sets of boundary conditions.

The results of the foregoing study may be summarized as follows:

- Despite the vertical boundaries of the porous slab being in relative motion, it is established that the base flow is stable for all values of the Prandtl-Darcy number $Pr_D$ and for all infinitesimal perturbations when the boundaries are impermeable-isothermal. In the limit $Pr_D \to \infty$, however, the proof of stability, as originally demonstrated by Gill, is shown to be applicable to the current problem.
- Under specific parametric conditions, the base flow undergoes instability when the impermeable boundaries are either partially conducting or perfectly adiabatic, regardless of the values of $Pr_D$. This instability is observed for both stationary and moving plates. This is another instance where Gill's theorem fails. The critical stability parameters, delineating the threshold for instability onset, are ascertained and noted that neutral stability conditions occur with traveling-wave modes. The initial range of $Bi$ within which flow remains stable increases as $Pr_D$ decreases, and a similar trend is observed for the $Pe$, although its influence remains relatively modest. The effect of $Pr_D$ on the stability of base flow is two-fold as it manifests both stabilizing and destabilizing tendencies depending upon the values of $Bi$, while $Pe$ promotes a destabilizing impact on the base flow. Notably, the adiabatic boundaries lead to a more rapid initiation of instability in comparison to partially conducting boundaries.
- The partially permeable or perfectly permeable isothermal boundaries initiate instability on the base flow contrary to the impermeable boundaries. In the limit $N_0 \to \infty$ and when $Pe = 0$, the results coincide with that laid out and solved by Barletta [8] using pressure-temperature formulation. As $N_0$ increases, the flow gets destabilized within a low to moderate range of values of $Pe$ but exhibits both stabilizing and destabilizing effects at higher values of $Pe$. The presence of Couette flow makes the system more stabilizing and it has no influence on the onset of instability in the limit $Pr_D \to \infty$. Besides, increasing $Pr_D$ shows a destabilizing effect on the basic flow. The most unstable perturbations are stationary when the walls are at rest while they become non-stationary when they are in motion.
- For some choices of governing parameters, two distinct regimes of $Bi$ exist linked to two disconnected travelling-wave neutral curves for the partially permeable and partially conducting boundaries. In the initial regime of $Bi$, the onset of instability shows weak dependence on both $Pr_D$ and $Pe$, whereas $Bi$ instills both stabilizing and destabilizing effects. In the subsequent regime of $Bi$, both $Pe$ and $Bi$ promote the occurrence of instability but the impact of $Pr_D$ is not consistent and can vary. However, increasing $N_0$ triggers the instability in both the regimes of $Bi$.

- When the boundaries are permeable and adiabatic, increasing $Pe$ has the effect of destabilizing the base flow, which becomes progressively weaker as $Pr_D$ assumes higher values.

Thus various combinations of temperature and velocity boundary conditions have resulted in a dramatic change in the stability of the basic flow in a vertical porous slab though the basic state remains the same for all the boundaries.

This paper presents several potential avenues for future development based on the obtained results. One direction of exploration involves investigating temporal instability from a non-modal perspective. This can be achieved by examining the time evolution of a perturbation wave packet expressed through a Fourier integral. This alternative approach to temporal analysis leads to the identification of an absolute instability condition, typically characterized as supercritical. Similarly, a non-modal framework could be considered for the study of spatial instability. By extending the analysis of spatial instability beyond the limitations of purely modal analysis and linear formulation, exciting research opportunities can arise in the field of fluid mechanics pertaining to fluid-saturated porous media.

**Declaration of interests**: The authors report no conflict of interest
**Data availability:** The data that support the findings of this study are available within the article.